\documentclass[aps,prd,floatfix,twocolumn,superscriptaddress,preprintnumbers,showpacs,longbibliography]{revtex4-2}
\usepackage[colorlinks,bookmarks=true,citecolor=blue,linkcolor=red,urlcolor=blue,citecolor=blue]{hyperref}
\usepackage{subfigure}
\usepackage{graphicx}% Include figure files
\usepackage{float,placeins} % plots
\usepackage{dcolumn}% Align table columns on decimal point
\usepackage{bm}% bold math
\usepackage{color}
\usepackage{amsmath, amssymb}
\usepackage{physics}
\usepackage{color}
\usepackage{tabularx}
\usepackage{qcircuit}
\usepackage{mathtools}
\usepackage[table]{xcolor}
\usepackage{multirow}
\usepackage{orcidlink}

\begin{document}

\title{Concurrent VQE for Simulating Excited States of the Schwinger Model}

\author{Yibin Guo \orcidlink{0000-0003-0435-1476}}
\affiliation{CQTA, Deutsches Elektronen-Synchrotron DESY, Platanenallee 6, 15738 Zeuthen, Germany}

\author{Takis Angelides \orcidlink{0000-0002-8639-8050}}
\affiliation{CQTA, Deutsches Elektronen-Synchrotron DESY, Platanenallee 6, 15738 Zeuthen, Germany}
\affiliation{Institut für Physik, Humboldt-Universität zu Berlin, Newtonstr. 15, 12489 Berlin, Germany}

\author{Karl Jansen \orcidlink{0000-0002-1574-7591}}
\affiliation{CQTA, Deutsches Elektronen-Synchrotron DESY, Platanenallee 6, 15738 Zeuthen, Germany}
\affiliation{Computation-Based Science and Technology Research Center, The Cyprus Institute, 20 Kavafi Street,
2121 Nicosia, Cyprus}

\author{Stefan K\"uhn \orcidlink{0000-0001-7693-350X}}
\affiliation{CQTA, Deutsches Elektronen-Synchrotron DESY, Platanenallee 6, 15738 Zeuthen, Germany}

\begin{abstract}
This work explores the application of the concurrent variational quantum eigensolver (cVQE) for computing excited states of the Schwinger model. By designing suitable ansatz circuits utilizing universal SO(4) or SO(8) qubit gates, we demonstrate how to efficiently obtain the lowest two, four, and eight eigenstates with one, two, and three ancillary qubits for both vanishing and non-vanishing background electric field cases. Simulating the resulting quantum circuits classically with tensor network techniques, we demonstrate the capability of our approach to compute the two lowest eigenstates of systems with up to $\mathcal{O}(100)$ qubits. Given that our method allows for measuring the low-lying spectrum precisely, we also present a novel technique for estimating the additive mass renormalization of the lattice based on the energy gap. As a proof-of-principle calculation, we prepare the ground and first-excited states with one ancillary and four physical qubits on quantum hardware, demonstrating the practicality of using the cVQE to simulate excited states. 
\end{abstract}

\maketitle

\section{Introduction\label{sec:intro}}

Investigating the energy spectrum and excited states is crucial in understanding various physical phenomena. The energy spectrum, especially the energy gap, encodes information for the phase transition in statistical physics~\cite{jaeger1998ehrenfest}, the transition energy in chemistry~\cite{serrano2005quantum}, and the electronic and optical properties in band theory~\cite{kittel2018introduction}. In high-energy physics, excited states correspond to different kinds of particles and bound states~\cite{gattringer2009quantum}. For instance, the low-energy excited states in Quantum Chromodynamics (QCD) correspond to the bound states of quarks with gluons, which can be directly observed and studied in experiments. This simple yet profound connection renders the study of excited states and energy spectra particularly valuable, making it possible to explain various experimental phenomena from first principles.

The Schwinger model~\cite{schwinger1962gauge} describes $1+1$-dimensional quantum electrodynamics (QED) and shares many properties with QCD, such as confinement, charge screening, U(1) chiral symmetry breaking, and a topological vacuum structure. 
Although the Schwinger model can be exactly solved in the limit of massless fermions~\cite{schwinger1962gauge}, the majority of parameter regimes requires numerical techniques for evaluation. Discretizing the model on a lattice, it can be addressed with several numerical techniques. Due to its similarities with QCD, it serves as an ideal testbed for new methods. Various classical numerical methods have been successfully applied to the Schwinger model~\cite{wilson1974confinement,banks1976strong,kogut1975hamiltonian,coleman1976more,adam1996schwinger}, which include exact diagonalization (ED)~\cite{hamer1997series,sriganesh2000new}, Markov chain Monte Carlo (MC)~\cite{marinari1981monte,duncan1981monte}, tensor networks (TN)~\cite{byrnes2002density,banuls2013mass,kuhn2014quantum,buyens2014matrix,rico2014tensor,buyens2016confinement,banuls2017density,buyens2017finite,zapp2017tensor,ercolessi2018phase,funcke2020topological,funcke2023exploring,pichler2016real,buyens2017real,sala2018variational,magnifico2020real,rigobello2021entanglement,shimizu2014grassmann,banuls2015thermal,banuls2016chiral,buyens2016hamiltonian,butt2020tensor,papaefstathiou2021density}, and other methods~\cite{banuls2020review}. However, these classical methods face some limitations. For ED, the exponential increase of the dimension of the Hilbert space limits the possible lattice sizes. For MC, the well-known sign problem occurs in certain parameter regimes, such as the presence of a topological $\theta$-term~\cite{funcke2020topological}. TN become intractable in most dynamical and high-dimensional systems because of the large entanglement~\cite{xiang2023density}. Hence, it would be highly desirable to have alternative lattice methods overcoming these limitations.

Quantum computers offer a promising avenue towards tackling these problems~\cite{banuls2020review,Banuls2020}. Although current noisy intermediate-scale quantum computers still suffer from a considerable level of noise~\cite{preskill2018quantum,bharti2022noisy}, quantum computers promise to generate highly entangled states with high fidelity~\cite{muschik2017u,farrell2023scalable} and to efficiently simulate out-of-equilibrium dynamics~\cite{mi2022time,kim2023evidence}. This potential has already been demonstrated for the Schwinger model using various quantum platforms, including superconducting circuits~\cite{klco2018quantum,de2022quantum,pomarico2023dynamical,farrell2023scalable,charles2024simulating}, trapped ions systems~\cite{martinez2016real,kokail2019self,nguyen2022digital,mueller2023quantum}, Rydberg atom systems~\cite{surace2020lattice}, ultra-cold atoms~\cite{mil2020scalable,yang2020observation,zhou2022thermalization,zhang2023observation}, and photonic systems~\cite{lu2019simulations}. 

So far, the majority of the quantum computing approaches to the Schwinger model has been focused on obtaining the ground state or on simulating dynamics. While there are several algorithms for computing excited states, in particular hybrid quantum-classical algorithms, which includes the orthogonality constrained VQE (OC-VQE)~\cite{higgott2019variational}, variational quantum deflation (VQD)~\cite{jones2019variational,kuroiwa2021penalty}, subspace search VQE (SS-VQE)~\cite{nakanishi2019subspace}, multistate contracted VQE (MC-VQE)~\cite{parrish2019quantum}, concurrent VQE (cVQE)~\cite{xu2023concurrent} and other methods~\cite{mcclean2017hybrid,santagati2018witnessing,ollitrault2020quantum,tilly2020computation,wen2024full,ding2024ground}, determining multiple excited states for lattice field theories on near-term quantum devices remains a challenge. In this work, we provide a scheme for obtaining multiple excited states of the lattice Schwinger model using cVQE, a variational algorithm based on the generalized Rayleigh-Ritz variational principle. We demonstrate that our cVQE approach allows for computing the scalar state of the theory and that it can be efficiently scaled up to large system sizes  with up to $\mathcal{O}(100)$ qubits.

The rest of the paper is organized as follows. Section.~\ref{sec:setup} introduces the Schwinger model and its lattice formulation with Kogut-Susskind staggered fermions. Subsequently, we will review cVQE and discuss the ansatz circuits used. Moreover, the relevant observables are introduced, as well as how to obtain them on quantum devices. After presenting the setup, we will report the results of the lowest two eigenstates with one ancillary qubit in Sec.~\ref{sec:One-ancilary}, both for small and large systems with qubit numbers up to $\mathcal{O}(100)$.  We then present results for the lowest eight eigenstates with three ancillary qubits in Sec.~\ref{sec:Three-ancilary}, showing that we can obtain the energy pseudo-momentum dispersion, allowing us to identify the vector and scalar branches. All the results considered in Sec.~\ref{sec:One-ancilary} and \ref{sec:Three-ancilary} include cases with vanishing and non-vanishing background electric field.
Subsequently, we will discuss the connection between the energy gap and the additive mass renormalization of the model in Sec.~\ref{sec:massshift}. We present a method that allows for measuring the mass shift by tuning the energy gap of the lattice model to the free Schwinger boson mass. Finally, we demonstrate the preparation of excited states on IBM's $ibm\_algiers$ device in Sec.~\ref{sec:experiment}.

\section{Model and methods\label{sec:setup}}

\subsection{The Schwinger model and its Kogut-Susskind Hamiltonian}
The Schwinger model describes quantum electrodynamics in $1+1$-dimensional space-time. 
The Lagrangian density of the massive Schwinger model with a topological $\theta$-term reads as
\begin{equation}
\label{eq:lagrangian_density}
\mathcal{L}=\bar{\Psi}\left(\mathrm{i}\gamma^{\mu}\partial_{\mu}-g\gamma^{\mu}A_{\mu}-m\,\right)\Psi-\frac{1}{4}F_{\mu\nu}F^{\mu\nu}+\frac{g\theta}{4\pi}\epsilon^{\mu\nu}F_{\mu\nu},
\end{equation}
where the electromagnetic tensor is given by
\begin{equation}
    F_{\mu\nu}=\partial_{\mu}A_{\nu}-\partial_{\nu}A_{\mu},
\end{equation}
with $g$ the coupling constant and $m$ the fermion mass. Here, $\Psi$ represents the fermionic field with a two-component spinor, and $g\theta/2\pi$ is the static background electric field~\cite{coleman1976more,banuls2013mass,funcke2020topological}.
Schwinger~\cite{schwinger1962gauge} originally solved this model in the massless limit, $m=0$, and provided an analytical expression for the energy gap (known as free Schwinger boson mass)
\begin{equation}
    \frac{M_S}{g}\bigg|_{m/g=0} = \frac{1}{\sqrt{\pi}}.
    \label{Eq:MassSchwingerBoson}
\end{equation}
Using mass perturbation theory, Adam~\cite{adam1996schwinger} derived the energy gap for small mass
\begin{align}
    \frac{M_S}{g} = &\frac{1}{\sqrt{\pi}} \Bigg[ 1 + 2 \sqrt{\pi} e^{\gamma} \left( \frac{m}{g} \right) \cos \left(\theta\right) \nonumber \\
    & + \pi e^{2\gamma} \left( \frac{m}{g} \right)^2 \big(-0.6599 \cdot \cos \left(2\theta\right) + 1.7277 \big) \Bigg]^{\frac{1}{2}}
    \label{Eq:MassSchwingerPerturbation}
\end{align}
to second order in $m/g$, where $\gamma \approx 0.5772156649$ is Euler's constant.

\begin{figure}[!htbp]
    \centering
    \includegraphics[width=0.35\textwidth]{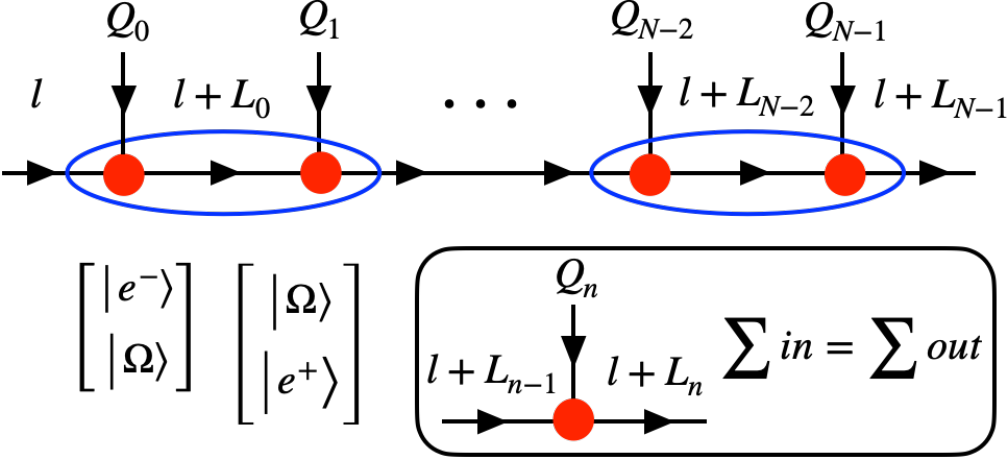}
    \caption{
     The Kogut-Susskind staggered formulation of the Schwinger model. The red circles are the lattice sites where the field $\phi_n$ is placed. The horizontal arrows is where the fields $\vartheta_n$, $L_n$ are placed and the vertical arrows and the direction of all arrows aid with demonstrating Gauss's law shown in the black box. We start counting sites from 0. For the even sites, spin up (down) represents the electron (vacuum $\ket{\Omega}$) state. For the odd sites, spin up (down) represents the vacuum (positron) state. The $Q_n$ is the total charge on site $n$. The no-source condition at each lattice site corresponds to Gauss law in Eq.~\eqref{Eq:gauss}.
     }
     \label{fig:staggerfermion}
\end{figure}

Using the temporal gauge where $A_0 = 0$, the corresponding Hamiltonian density of Eq.~\eqref{eq:lagrangian_density} reads
\begin{equation}
    \mathcal{H}=-\mathrm{i}\bar{\Psi}\gamma^{1}\left(\partial_{1} - \mathrm{i} gA_{1}\right)\Psi+m\bar{\Psi}\Psi+\frac{1}{2}E^{2}
    \label{eq:Hamiltonian_continuum}
\end{equation}
with the Gauss law constraint
\begin{equation}
    \partial_{1} E = g \bar{\Psi}\gamma^{0} \Psi.
\end{equation}
We discretize Eq.~\eqref{eq:Hamiltonian_continuum} on a lattice and make use of the Kogut-Susskind staggered formulation~\cite{kogut1975hamiltonian}.
The lattice Hamiltonian then reads as
\begin{align}
    H =& -\frac{\mathrm{i}}{2a} \sum_{n=0}^{N-2} \left(\phi_{n}^{\dagger} e^{\mathrm{i}\vartheta_{n}}\phi_{n+1} - \phi_{n+1}^{\dagger} e^{\mathrm{-i}\vartheta_{n}}\phi_{n}\right) \\ \notag
    & + m_{\mathrm{lat}}\sum_{n=0}^{N-1} \left( -1 \right)^{n} \phi_{n}^{\dagger}\phi_{n} + \frac{a g^{2}}{2} \sum_{n=0}^{N-2} (l+L_{n})^{2}.
\end{align}
Here, $m_{\mathrm{lat}}$ is the lattice mass, $a$ is the lattice spacing and $\vartheta_{n}$ is the gauge variable defined by $\vartheta_{n}=-agA_{n}^{1}$. The field $\phi_{n}$ is a single-component fermion field at site $n$. The staggered charge operator on the lattice is given by
\begin{equation}
    Q_n = \phi_n^{\dagger}\phi_n -\frac{1}{2}\left[1-\left(-1\right)^{n}\right].
\end{equation}
The operators $L_{n}$ and $\vartheta_{n}$ act on the link to the right of site $n$, and are conjugate variables satisfying the condition
\begin{equation}
    \left[\vartheta_{n},L_{m}\right]=i\delta_{nm},
\end{equation}
with $\delta_{nm}$ the Kronecker delta. $L_{n}$ is related to the electric field by $g L_{n}=E_{n}$. The constant background electric field $l={\theta}/{2\pi}$ corresponds to a topological term with angle $\theta$. The physical states of the Hamiltonian have to fulfill Gauss law
\begin{equation}
    L_{n}-L_{n-1} = Q_n,
    \label{Eq:gauss}
\end{equation}
which corresponds to the no source condition shown in the black box of Fig.~\ref{fig:staggerfermion}.

To address the theory with a quantum computer, we have to map the Hamiltonian to qubits. To this end, we use the Jordan-Wigner transformation
\begin{equation}
    \phi_{n} = \Bigg[\prod_{k<n} \left( \mathrm{i} \sigma_k^z \right) \Bigg] \left(\sigma_n^x - \mathrm{i} \sigma_n^y \right)/2,
\end{equation}
which allows us to map the fermionic fields to spins. The staggered charge in spin formulation reads
\begin{equation}
    Q_n = \frac{1}{2}\big[\sigma_n^z + (-1)^n \big].
\end{equation}

For our lattice simulations, it is convenient to work with a dimensionless version of the Hamiltonian $W=2H/ag^2$.
Here we focus on the case of open boundary conditions, which allows for integrating out the gauge fields and obtaining a Hamiltonian directly constrained to the gauge invariant subspace~\cite{hamer1997series,banuls2013mass}
\begin{align}
    W =& x\sum_{n=0}^{N-2}\left[\sigma_{n}^{+}\sigma_{n+1}^{-}+\sigma_{n}^{-}\sigma_{n+1}^{+}\right]+\frac{\mu}{2}\sum_{n=0}^{N-1}\left[1+\left(-1\right)^{n}\sigma_{n}^{z}\right] \nonumber \\
    & +\sum_{n=0}^{N-2}\left[l+\frac{1}{2}\sum_{k=0}^{n}\left(\left(-1\right)^{k}+\sigma_{k}^{z}\right)\right]^{2}. 
    \label{Eq:SpinHamiltonian}
\end{align}
In the expression above we have defined the quantities $x\equiv 1/g^{2}a^{2}$, $\mu\equiv2m_{\mathrm{lat}}/g^{2}a$ and added a constant term $\mu N/2$ to the Hamiltonian to set the ground state energy in the limit $m_{\mathrm{lat}}/g\rightarrow \infty$ to $0$~\cite{hamer1997series}. For the rest of the paper, we focus solely on the states with zero total charge, which in spin formulation is identical to the condition $\sum_n \sigma_n^z = 0$.
In simulations, we consider the positive semidefinite penalty term
\begin{equation}
    W_p = \lambda\left(\sum_{n=0}^{N-1}\sigma_{n}^{z}\right)^{2}
\label{Eq:PenaltyHamiltonian}
\end{equation}
with $\lambda>0$ to ensure we obtain states with vanishing total charge. 

\subsection{Concurrent variational quantum eigensolver}
\label{sec:cvqe} 
This subsection briefly revisits the cVQE which is a hybrid quantum-classical algorithm designed to determine the low-lying eigenstates, for more details see~\cite{xu2023concurrent}. cVQE is based on the generalized Rayleigh-Ritz variational principle using the following cost function 
\begin{equation}
    \bar{E}(\{ \theta_i \}) =  \frac{1}{\sqrt{K}}\sum_{m=0}^{K-1} \langle \psi_m(\{\theta_i\})|H|\psi_m(\{\theta_i\}) \rangle,
    \label{eq:cVQE_cost_function}
\end{equation}
where $|\psi_m(\{\theta_i\}) \rangle = U(\{\theta_i\}) |\psi_m^0 \rangle$, $U(\{\theta_i\})$ is a parametric unitary and $\{|\psi_m^0 \rangle\}$ are a set of orthogonal initial states. Note that in Eq.~\eqref{eq:cVQE_cost_function} the weights for the contributions of the individual eigenstates are the same, $1/\sqrt{K}$, to optimize all target states equally.
To facilitate the simultaneous simulation and efficient post-processing of various excited states, cVQE utilizes the purification technique to integrate all orthogonal initial states in one circuit and thereby enabling the reuse of conventional VQE codes. By maximally entangling the initial states $| \psi_m^0\rangle$ with the orthogonal states $| \alpha_m \rangle$ defined on ancillary qubits, we get the purified initial state~\cite{verstraete2004matrix}
\begin{equation}
    | \tilde{\Psi}_0 \rangle = \frac{1}{\sqrt{K}} \sum_{m=0}^{K-1} | \psi_m^0 \rangle \otimes | \alpha_m \rangle.
\end{equation}
The cost function can then be expressed as
\begin{equation}
   \bar{E}(\{ \theta_i \}) = \langle\tilde{\Psi}_0| U^{\dagger}(\{\theta_i\}) H U(\{\theta_i\}) |\tilde{\Psi}_0 \rangle,
   \label{Eq:costfunction}
\end{equation}
where the parameterized unitaries are applied only to the physical qubits. Thus, the cost function can be measured on quantum devices in a standard manner after decomposing the Hamiltonian in Pauli strings. 

\begin{figure}[!htbp]
    \centering
    \includegraphics[width=0.4\textwidth]{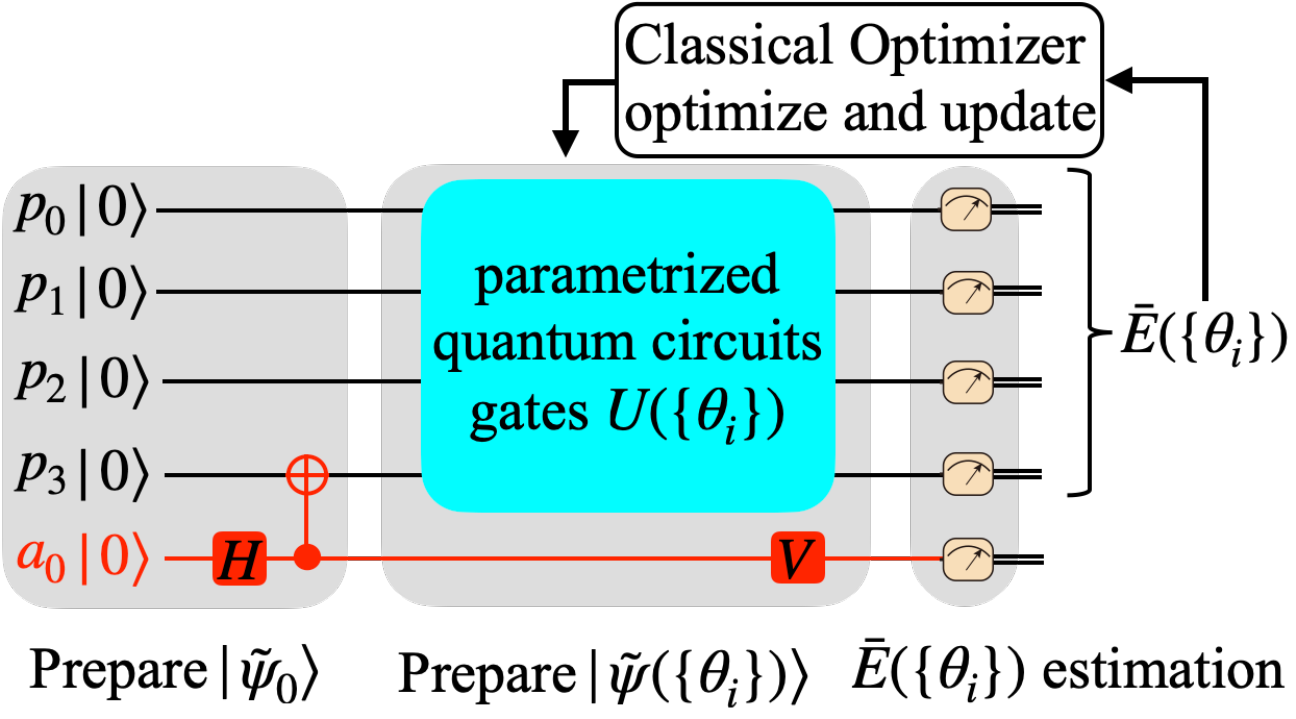}
    \caption{
    The concurrent variational quantum eigensolver (cVQE), which is a hybrid quantum-classical algorithm designed to find the low-energy eigenstates based on the generalized Rayleigh-Ritz variational principle. The ancillary qubit is introduced to prepare the maximally entangled state $|\tilde{\Psi}_0\rangle$ for the application of the purification technique. The parameterized gates $U(\{\theta_i\})$ are only applied on the physical qubits to prepare the state $|\tilde{\Psi}(\{\theta_i\})\rangle$. Several observables on quantum states are measured to estimate the energy expectation value $\bar{E}(\{\theta_i\})$. The classical optimizer is used to update the parameters by minimizing the $\bar{E}(\{\theta_i\})$. The loss function is invariant under the rotation given by $V$, hence the application of $V$ is only required at the end of the whole optimization to obtain the lowest eigenstates.
    }  
    \label{fig:cVQE}
\end{figure}

After minimizing the cost function, an additional rotation $V$ on the ancillary qubits is applied to obtain the true eigenstates $\{|E_m(\{\theta^{*}_i\}) \rangle\}$ from the optimized orthonormal states $\{|\psi_m(\{\theta^{*}_i\}) \rangle\}$. This rotation $V$ diagonalizes the Hamiltonian within the subspace spanned by $\{|\psi_m(\{\theta^{*}_i\}) \rangle\}$. The maximally entangled nature of the ancillary qubits with the physical states allows for determining all diagonal and off-diagonal matrix elements, $H_{mn} = \langle \psi_m(\{\theta^{*}_i\}) | H |\psi_n(\{\theta^{*}_i\}) \rangle$, by measuring the expectation values of $H$ and operators on ancillas. As an example, we give the whole procedure of cVQE with four physical qubits and one ancillary qubit in Fig.~\ref{fig:cVQE}.

\subsection{Circuit structures}
In general, the ansatz circuit of cVQE comprises three elements, including the location of the physical qubits maximally entangled with the ancillary qubits, the choice of the parameterized quantum gates, and the circuit structure. This subsection will introduce the ansatz circuit used in this paper by discussing these three elements in detail.

\begin{figure*}[!htbp]
    \centering
    \includegraphics[width=0.95\textwidth]{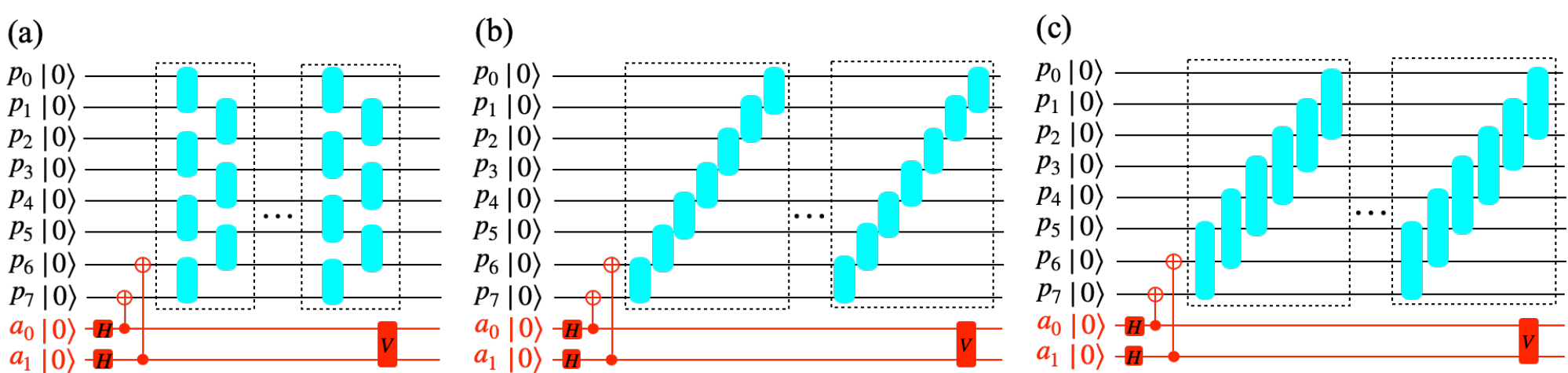}
    \caption{
    The circuit structure. The qubits $a_{n}$ ($p_{n}$) colored in red (black) represent the ancillary (physical) qubits. The Hadamard gate together with the CNOT gate generate a maximally entangled pair. The universal SO(4) gates (two-qubit gates colored in cyan) are arranged with brick-wall structure in (a) and ladder structure in (b). The universal SO(8) gates (three-qubit gates colored in cyan) are arranged with ladder structure in (c).
    All qubit gates in one dashed box form a qubit gate layer.
    }
    \label{fig:circuit_structure}
\end{figure*}

Let us start by discussing the distribution of the maximally entangled pairs. As illustrated in Fig.~\ref{fig:circuit_structure}, the ancillary qubits, denoted as $a_{n}$ and colored in red, are distinguished from the physical qubits, represented as $p_{n}$ colored in black. Acting with Hadamard gates on the ancillas, and subsequently applying a CNOT between the ancillas and the physical qubits, will generate a state with maximally entangled pairs $|\tilde{\psi}_0\rangle$. Physical qubits are sequentially entangled with ancillary qubits, starting from one end of the arrangement. This strategy circumvents long-range qubit gates for preparing initial states within a one-dimensional setup. The locations of these maximally entangled pairs can also be adapted, allowing for customization in accordance with the topology of existing quantum hardware platforms. 

\begin{figure}[!htbp]
    \centering
    \includegraphics[width=0.4\textwidth]{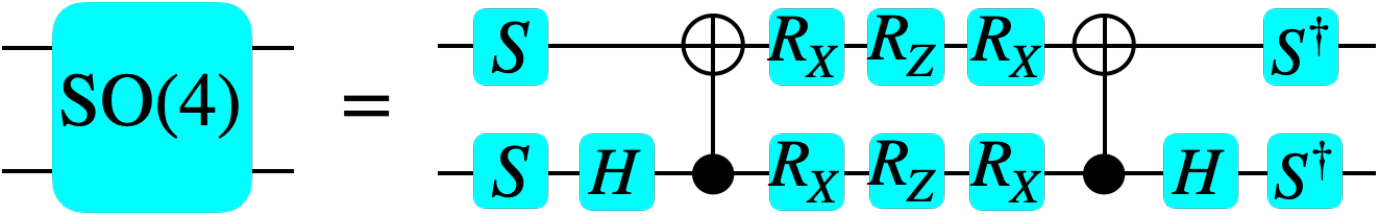}
    \caption{
     The decomposition of a universal SO(4) gate into two CNOT gates, two universal single-qubit gates, four phase gates, and two Hadamard gates. Each universal single-qubit gate can be further decomposed into parameterized gates $R_X, R_Z, R_X$, where $R_Q(\theta)$ is a rotation around $Q$-axis and $R_Q(\theta) = \mathrm{exp}(-\mathrm{i} \theta Q/2)$.
     }
     \label{fig:gate_decomposition_so4}
\end{figure}

In our simulation, we observed that the universal SO(4) gates have enough expressive ability to accurately simulate the lowest eigenstates when employing one or two ancillary qubits, which allows for computing up to 4 eigenstates. However, when more than two ancillary qubits are introduced, we find that using SO(8) gates becomes more effective for computing the lowest eigenstates. Using SO(4) gates for more than two ancillary qubits, one observes that some eigenstates may be skipped, a phenomenon also found in Ref.~\cite{xu2023concurrent}. In practical applications, it is convenient to construct SO(4) gates or SO(8) gates $U$ as an exponential of a strictly triangular matrix $A$ according to
\begin{equation}
    U = \mathbf{exp}(A-A^T).
    \label{Eq:so(n)exp}
\end{equation}
The resulting unitary gates $U$ can be decomposed in elementary single- and two-qubit gates. As demonstrated in Ref.~\cite{vatan2004optimal}, a universal SO(4) gate can be decomposed into two CNOT gates, two universal parameterized single-qubit gates, and some single-qubit gates without parameters (see Fig.~\ref{fig:gate_decomposition_so4} for an illustration). 

Lastly, we discuss the circuit structure used in our numerical simulations. For NISQ devices, it is crucial to guarantee sufficient expressive ability with the limited number of entangling layers. Given this constraint, the universal SO(4) two-qubit gates, configured in a brick-wall pattern as illustrated in Fig.~\ref{fig:circuit_structure}(a), have been found to be adequate for capturing up to eight lowest eigenstates in smaller systems, for example, $8$ physical qubits systems with $3$ ancillary qubits. For larger systems with one ancillary qubit, the universal SO(4) two-qubit gates arranged in a ladder structure, as depicted in Fig.~\ref{fig:circuit_structure}(b), were found sufficient to simulate the two lowest eigenstates. For larger systems where the number of ancillary qubits exceeds two, the universal SO(8) three-qubit gates, arranged in a ladder structure as shown in Fig.~\ref{fig:circuit_structure}(c), were confirmed to be adequate for simulating the lowest eigenstates.

\subsection{Observables\label{subsec:quantitiesdef}}
Besides the fidelity $F$ of the states obtained from the cVQE with a reference state and the variance of the Hamiltonian, we monitor the observables described below to assess if we can reproduce the correct low-energy spectrum of the Hamiltonian.

\textbf{Quasimomentum}. The masses in a quantum field theory essentially correspond to the energy difference between the ground state and zero-momentum excitations with certain quantum numbers, where the momentum operator for a fermion field $\Psi(x)$ is well defined as $P = \mathrm{i} \int \mathrm{d}x \Psi^{\dagger}(x)\partial_x \Psi(x)$. The dimensionless staggered lattice discretization of this operator in spin form reads~\cite{banuls2013mass}
\begin{equation}
    O_{p}=-ix\sum_{n}\left[\sigma_{n}^{-}\sigma_{n+1}^{z}\sigma_{n+2}^{+}-\sigma_{n}^{+}\sigma_{n+1}^{z}\sigma_{n+2}^{-}\right].
\end{equation}
For a finite system with open boundary conditions there is no exact translational symmetry, and hence no proper definition of a pseudomomentum. However, as demonstrated in Ref.~\cite{banuls2013mass}, the expected value of $O_{p}^2$ still allows for a reliable construction of the dispersion and the identification of the zero-momentum excitations.

\textbf{Spin Transformation ($S_R$)}. In the context of the Schwinger model, the charge conjugation quantum number classifies excited states into vector ($C=-1$) and scalar ($C=+1$) states~\cite{hamer1997series,banuls2013mass}. In the staggered formulation for periodic boundary conditions, charge conjugation corresponds to a shift by one lattice site followed by a spin flip
\begin{equation}
   S_R = \bigotimes_{k=1}^{N/2} \sigma^x_{2k-1}T^{(1)},
\end{equation}
where $T^{(1)}$ denotes the translation by one spin site. This transformation no longer directly describes charge conjugation for open boundary conditions. However, as pointed out in Ref.~\cite{banuls2013mass}, the phase of the expectation value of $S_R$ still retains the information of the charge conjugation. A phase $\phi(\langle S_R \rangle) \approx 0$ indicates scalar states, while $\phi(\langle S_R \rangle) \approx \pi$ indicates vector states. This study utilizes this method to discern between different branches of excited states.

\textbf{Chiral condensate}. The chiral condensate serves as the order parameter for the chiral symmetry-breaking phase in the Schwinger model. In the continuum limit, the chiral condensate $\Sigma$ is the expectation value of the operator $\hat{\Sigma} = \bar{\Psi} \Psi$. For the staggered lattice discretization and applying the Jordan-Wigner transformation the corresponding operator is given by
\begin{equation}
    \hat{\Sigma} = \frac{g\sqrt{x}}{2N} \sum_{n=0}^{N-1} (-1)^n (1 + \sigma_n^z).
\end{equation}
In the massless limit, $\Sigma$ can be derived analytically~\cite{sachs2010finite}. However, analytical solutions are unavailable in the massive regime, necessitating numerical approaches for precise determination.

\textbf{Electric field density (EFD)}. To reduce finite-size effects in the lattice model with open boundary conditions, we take the average of the electric field on the $2r$ gauge links in the center of the system to measure the EFD. The EFD can then be expressed as
\begin{equation}
    F_{av}=\frac{1}{2r}\sum_{k=0}^{r-1}\left(L_{\frac{N}{2}-k-1}+L_{\frac{N}{2}+k}\right).
\end{equation}
For nonzero $l$, the induced fermion/anti-fermion pairs screen the background electric field, where complete screening occurs in the massless limit. This complete screening effect can be used to determine the additive mass renormalization as outlined in the the next paragraph.

\textbf{Mass shift}. The staggered discretization leads to an additive mass renormalization~\cite{Angelides2022,dempsey2022discrete,angelides2023computing}. The renormalized mass $m/g$ can be expressed as the lattice mass $m_{\mathrm{lat}}/g$ plus an additive shift $m_s(x,N,l)/g$
\begin{equation}
    \frac{m}{g} = \frac{m_{\mathrm{lat}}}{g} + \frac{m_s(x,N,l)}{g},
    \label{eq:mass_shift}
\end{equation}
where the shift depends on the parameters $x$, $N$, and $l$, as shown in Ref.~\cite{Angelides2022,angelides2023computing,dempsey2022discrete}. For $l\neq 0$, one can identify the additive mass renormalization from the condition $F_{av}=0$ which corresponds to $m/g=0$~\cite{Angelides2022,angelides2023computing}. As we demonstrate in Sec.~\ref{sec:massshift}, the additive shift can also be extracted from the gap of the theory for arbitrary values of $l$.

\section{The lowest two eigenstates with one ancillary qubit\label{sec:One-ancilary}}
Let us begin with the case of one ancillary qubit, which allows for computing one excited state. To assess the performance of cVQE, we simulate it classically assuming a perfect, noise-free quantum device, where we proceed in two steps. First, we consider the performance of cVQE for small-scale systems with 8 qubits for both vanishing and nonvanishing background electric field. In these cases, the system can be easily treated with ED and we can compare the cVQE data to the exact results. Second, we show that our approach can be scaled up to $\mathcal{O}(100)$ qubits, thus allowing to go beyond system sizes accessible with ED. In this case, we use TN techniques to simulate the cVQE and compare the results with those from matrix product states (MPS). The quasi-Newton L-BFGS algorithm~\cite{liu1989limited} is adopted to minimize the loss function Eq.~\eqref{Eq:costfunction} in all cVQE simulations.

\subsection{Vanishing Background Electric Field}
\begin{figure}[!htb]
    \centering
    \includegraphics[width=0.45\textwidth]{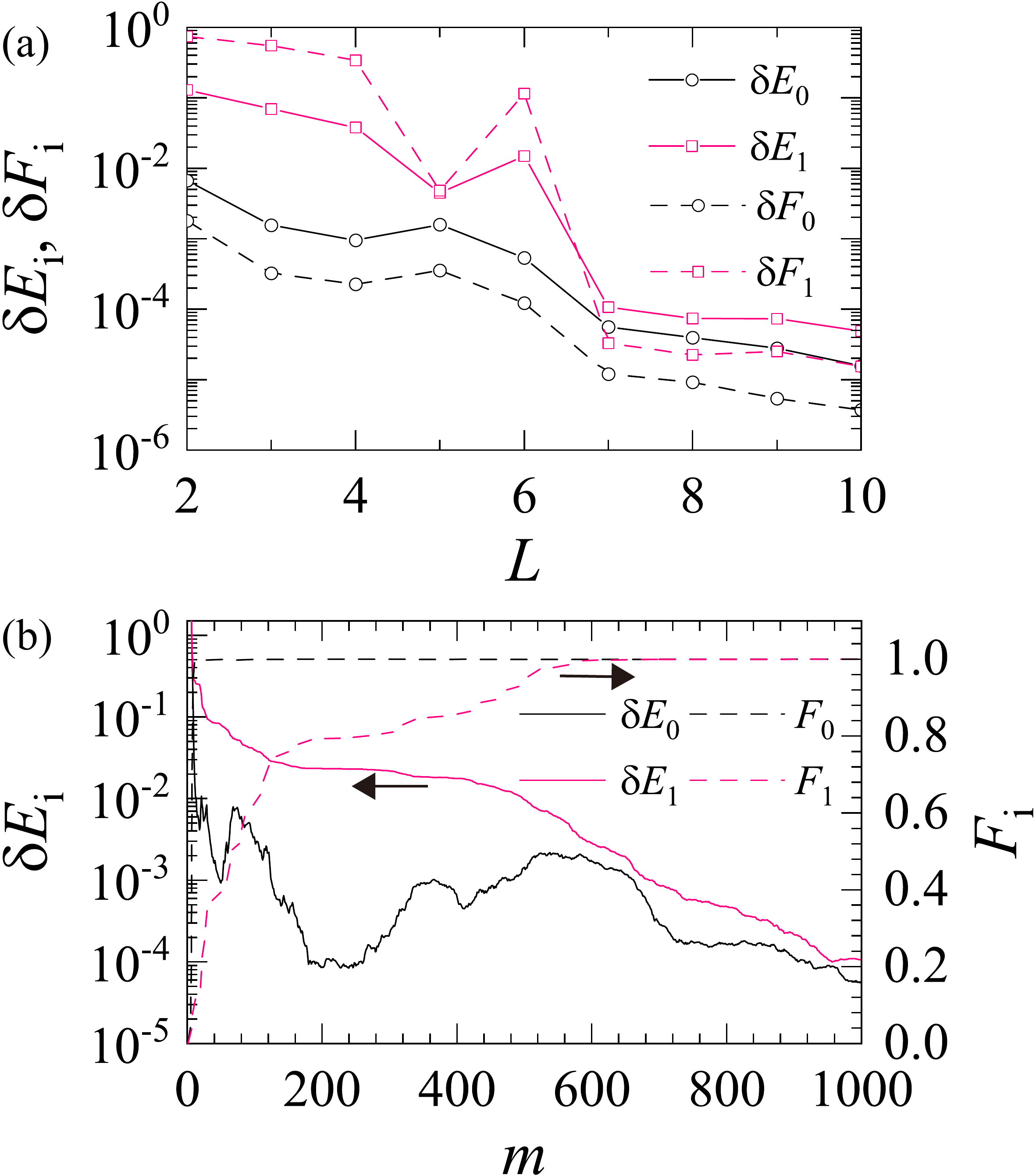}
    \caption{
    Convergence of the energy and the fidelity as a function of the number of layers (a) and as a function of the number of iterations for $L=7$ (b) for $N/\sqrt{x}=10$, $x=0.64$, $m_{\mathrm{lat}}/g=0.125$ and $l=0$. The black solid lines and data points correspond to the results for the ground state, the pink squares are the data for the first excited state. In panel (b) the $x$-axis shows the iteration, the left (right) $y$-axis is the energy difference (fidelity) between cVQE results and ED, represented as the solid (dashed) line. The results are selected from eleven random seeds of initial parameters according to the lowest expectation value of Eq.~\eqref{Eq:SpinHamiltonian} with the constraint in Eq.~\eqref{Eq:PenaltyHamiltonian}.
    }
    \label{fig:N8A1bf0mg125}
\end{figure}

As a first benchmark, we consider a small-scale system with $N=8$ physical qubits without background electric field, i.e.\ $l=0$. We choose a dimensionless lattice volume $N/\sqrt{x} = 10$, which corresponds to $x=0.64$, and set $m_{\mathrm{lat}}/g$ to $0.125$. To enforce vanishing total charge, we use the penalty term from Eq.~\eqref{Eq:PenaltyHamiltonian} with $\lambda = N$. For this small-size system with one ancillary qubit, we adopt the circuit ansatz shown in Fig.~\ref{fig:circuit_structure}(a). First, we study how the precision in energy, $\delta E_i = E_i - E_i^{ed}$, and the infidelity, $\delta F_i = F_i - F_i^{ed}$, for the $i$-th state obtained from cVQE compared to the ED results converge with the number of layers. Figure~\ref{fig:N8A1bf0mg125}(a) shows the data for $L=2$ to $10$. We observe that increasing the number of layers initially leads to an increase in precision before approximately reaching a plateau close to $L=7$ around $10^{-4}$ for $\delta E_i$. Increasing the number of layers beyond $L=7$ does not significantly enhance the precision, despite adding more parameters to the ansatz.  

Focusing on $L=7$, we show the typical convergence as a function of the iterations of the cVQE in Fig.~\ref{fig:circuit_structure}(b). The $x$-axis is the optimization step, and the left (right) $y$-axis is the energy difference (fidelity). While we observe quick convergence for the ground state after just a few tens of iterations, the first excited state requires more optimization steps. To reach a $\delta E_i$ of $10^{-4}$ for both states, we require around 1000 steps, where we observe final fidelities close to 1 ($0.99998796$ for the ground state and $0.99996709$ for the first excited state). After 1000 optimization steps, the variance of the Hamiltonian of the ground (first excited) state is $0.00030249$ ($0.00046949$), indicating that the simulation states are close to the reference eigenstates. In addition, we observe for both states $|\langle \sum_i \sigma_i^z \rangle| < 10^{-6}$ showing that they have essentially vanishing total charge. These results demonstrate that the cVQE with the circuit ansatz shown in Fig.~\ref{fig:circuit_structure}(a) has enough expressivity to capture the lowest two eigenstates and, therefore, the energy gap to high precision for small-size systems with vanishing background electric field.

\subsection{Non-vanishing Background Electric Field}
Next we turn to the case of non-vanishing background electric field, a regime where conventional MC methods suffer from the sign problem. We use the same parameters as before, except for setting $l=0.125$. To compare the results with the former vanishing background electric field case, we utilize the same brick-wall circuit ansatz with universal SO(4) gates of Fig.~\ref{fig:circuit_structure}(a) to determine the lowest two eigenstates for 8 physical qubits.

\begin{figure}[!htb]
    \centering
    \includegraphics[width=0.45\textwidth]{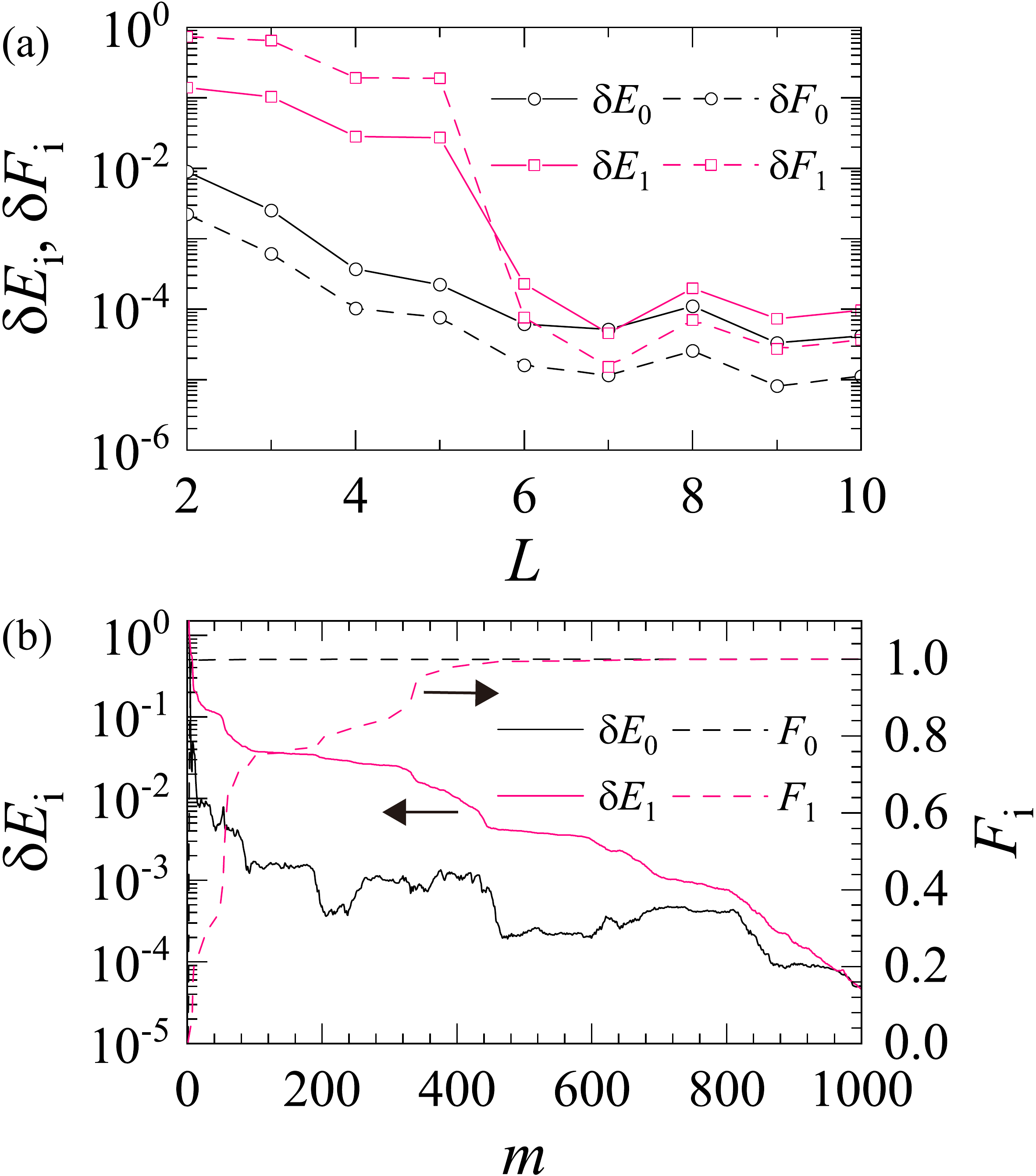}
    \caption{
    Convergence of the energy and the fidelity as a function of the number of layers (a) and as a function of the number of iterations for $L=7$ (b) for $N=8$, $x=0.64$, $m_{\mathrm{lat}}/g=0.125$ and $l=0.125$. The black solid lines and data points correspond to the results for the ground state, the pink squares are the data for the first excited state. In panel (b) the $x$-axis shows the optimization step, the left (right) $y$-axis is the energy difference (fidelity) between cVQE results and ED, represented as the solid (dashed) line. The results are selected from eleven random seeds of initial parameters according to the lowest expectation value of Eq.~\eqref{Eq:SpinHamiltonian} with the constraint in Eq.~\eqref{Eq:PenaltyHamiltonian}.
    }
    \label{fig:EFN8A1l0.125m0.125}
\end{figure}

In Fig.~\ref{fig:EFN8A1l0.125m0.125}(a), we show again the deviation in energy and fidelity from the ED results as a function of $L$. Compared to the case of vanishing background field in Fig.~\ref{fig:N8A1bf0mg125}(a), we observe that the background field does not lead to any qualitative differences. Both the values for $\delta E_i$ and $\delta F_i$ decrease to around $10^{-4}$ when $L \geq 6$. 
Again the typical convergence for 7 layers is illustrated in Fig.~\ref{fig:EFN8A1l0.125m0.125}(b).
The value of $\delta E_i$ decreases to around $10^{-2}$ after 400 steps of optimization and finally to around $10^{-4}$ after 1000 steps. The fidelity $F_i$ shows a similar behaviour as $\delta E_i$, it first increases to a value of approximately $0.99$ after 400 optimization steps and finally reaches $0.99998850$ ($0.99998489$) for the ground (first excited) state.  The variance of the Hamiltonian of the ground (first excited) state is $0.00025297$ ($0.00020323$) after 1000 optimization steps. For the penalty term from Eq.~\eqref{Eq:PenaltyHamiltonian}, we have $|\langle \sum_i \sigma_i^z \rangle| < 10^{-6}$ for both states. The small variance indicates that the simulated states are close to the real eigenstates. Our results demonstrate that the cVQE with the brick ansatz (see Fig.~\ref{fig:circuit_structure}(a)) is able to resolve the low lying spectrum with high accuracy, and implies that the energy gap can be determined to high precision for both vanishing and non-vanishing background electric field.

We also observe a similar behaviour if we add an additional ancillary qubit. The two ancillary qubits allow us to compute four eigenstates (see Appendix~\ref{sec:Two-ancilary} for details.)

\subsection{Simulating large systems\label{subsec:cVQElarge}}
Our small scale-results demonstrate that cVQE has the potential to determine excited states with high precision. An important question is if this success can also be achieved for system sizes that are beyond the capabilities of ED. Therefore, in this subsection, we study the behaviour of cVQE for computing excited states, and especially the energy gap, for system with up to $\mathcal{O}(100)$ sites. In our simulation, we find that the circuit structure in Fig.~\ref{fig:circuit_structure}(a) requires many more layers for large system sizes than the one in Fig.~\ref{fig:circuit_structure}(b) to get equally accurate results. The more layers in the ansatz, the more parameters the circuit has, which in turn leads to a more difficult optimization problem. To ensure sufficient circuit expressivity for large systems and an efficient optimization of the parameters, we will adopt the ansatz circuit from Fig.~\ref{fig:circuit_structure}(b) throughout this subsection.

To determine the typical precision of the energy gap obtained by cVQE, we add an ancillary qubit, allowing us to determine the lowest two eigenstates. Similar to the small-scale case, the penalty term from Eq.~\eqref{Eq:PenaltyHamiltonian} is added to the Hamiltonian to enforce vanishing total charge. In our numerical calculations, we find that 8 layers of the ansatz in Fig.~\ref{fig:circuit_structure}(b) are enough for the cVQE to produce high fidelity results ($F_i$ around $0.99$). Just as for 8 qubits, we also focus here on a dimensionless volume of $N/\sqrt{x}=10$ for vanishing and non-vanishing background field. For $l=0$, we study both vanishing lattice mass and $m_{\mathrm{lat}}/g=0.125$, while for a nonvanishing background field of $l=0.125$, we restrict ourselves to $m_{\mathrm{lat}}/g=0.125$ without loss of generality. To benchmark the performance of cVQE, we utilize matrix product states (MPS) with bond dimension $D = 40$ as the reference states (details on the MPS methods are given in Appendix~\ref{app:mps}).

The optimization process in the cVQE is divided into two steps. First, we perform $500$ iterations and optimize the gate parameters with the constraint that the gates have translational symmetry. This allows for reducing the number of parameters in the optimization, and for large enough system sizes the physical states are still expected to be translation invariant in the bulk region even for open boundary conditions. At this stage, we set the $\lambda$ of the penalty term Eq.~\eqref{Eq:PenaltyHamiltonian} to $0.5$ per site for system sizes $N=20, 40$ and to $0.4$ per site for system sizes $N=60, 80, 100$. Second, the translation invariance constraint is removed, which allows for incorporating the effects induced by open boundary conditions while having a good starting point. All gates are then further optimized independently with an additional $2000$ to $4000$ iterations, depending on the convergence behaviour. At this stage, we also decrease the $\lambda$ by a factor of 2 to in general reduce the weight of the penalty term on the total energy. We find that this choice ensures $|\langle \sum_i \sigma_i^z \rangle| < 10^{-3}$ in all our simulations. After obtaining the optimized SO(4) gates, these can be decomposed into basic gate operations as shown in Fig.~\ref{fig:gate_decomposition_so4}.

\begin{figure}[!htp]
    \centering
    \includegraphics[width=0.45\textwidth]{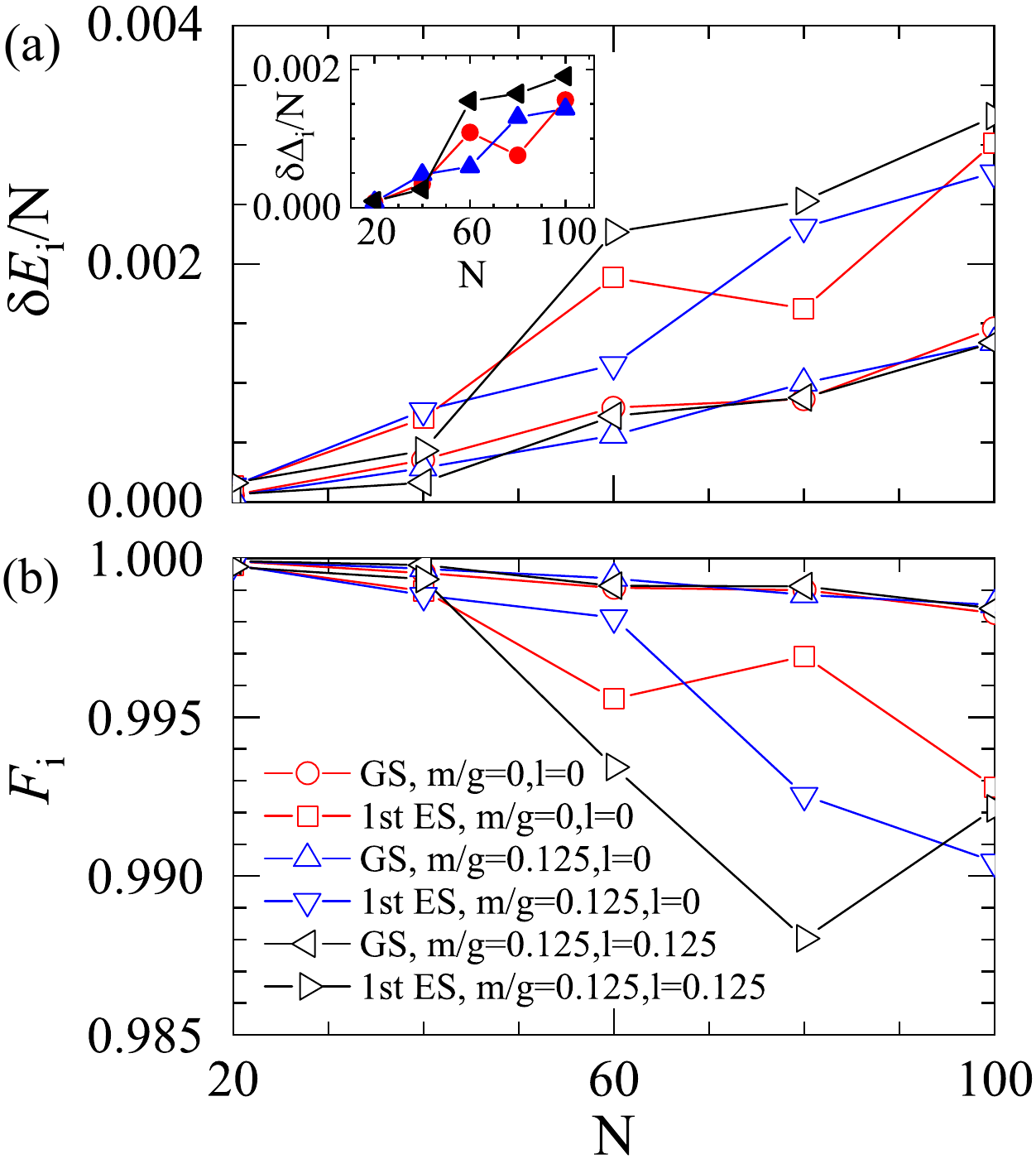}
    \caption{
      Energy and fidelity behaviour of the ground state and the first excited state of different qubit numbers using the circuit ansatz \ref{fig:circuit_structure}(b). Here we focus on the parameter region of volume 10. The qubit gates layers here is 8. (a) The per site energy difference $\delta E_i/N$ between the results of cVQE and MPS (D = 40). The inset shows how the energy gap per site difference ($\delta \Delta_i / N $) between the cVQE and MPS of the i-th eigenstate changes with system size N. The red filled circle (blue filled triangle or black filled left triangle) represents $m_{\mathrm{lat}}/g = 0$ and $l = 0$ ($m_{\mathrm{lat}}/g = 0.125$ and $l = 0$ or $m_{\mathrm{lat}}/g = 0.125$ and $l = 0.125$). (b) The fidelity between states from cVQE and MPS.
     }
     \label{fig:ScalingcVQEGap}
\end{figure}

Figure~\ref{fig:ScalingcVQEGap} shows the results obtained from cVQE as a function of system size, where the lowest energy results obtained from 11 random initial parameters are depicted. 
Focusing on the energy difference between cVQE and MPS for the $i$-th eigenstate, $\delta E_i$, in Fig.~\ref{fig:ScalingcVQEGap}(a), we observe that both $\delta E_0/N$ and $\delta E_1/N$ show an approximately linear increase with $N$. One important reason for this phenomenon is the increase of the state's energy with the system size. One can apply more ansatz layers to get better results in larger systems if the same $\delta E_i$ is desired. All $\delta E_i$ reach at least $\mathcal{O}(10^{-1})$, while the errors in energy per site are on the order of $ |\delta E_i / N| \thicksim \mathcal{O}(10^{-3})$.
The inset in Fig.~\ref{fig:ScalingcVQEGap}(a) shows the behaviour of the energy gap density difference $\delta \Delta_i/N$ between cVQE and MPS for $i$-th eigenstate, where the red filled circles (blue filled triangles / black filled left triangles) represent the case of $m_{\mathrm{lat}}/g = 0$ and $l = 0$ ($m_{\mathrm{lat}}/g = 0.125$ and $l = 0$, $m_{\mathrm{lat}}/g = 0.125$ and $l = 0.125$). All the differences in the energy gap per site between the cVQE and the MPS calculation $\delta \Delta_i/N$ show excellent performance, and the relative error $\delta \Delta_i / N $ are smaller than $0.002$. 

Figure~\ref{fig:ScalingcVQEGap}(b) presents the fidelity $F_i$ between $i$-th state obtained from the cVQE and the corresponding state from the MPS calculation. Keeping the number of layers fixed at 8 for all the system sizes we study, we can reach fidelities of greater than $98.7\%$ throughout the entire range of system sizes we study. There is only a slight decrease from $0.999$ ($0.999$) for the ground state (first excited state) having $N=20$, to around $0.998$ ($0.990$) having $N=100$ for all of mass and background field values we consider. Moreover, we generally observe higher fidelity for the ground states compared to the excited states for all parameters sets we study. The observed behaviour of fidelity is consistent with the data for the energy. Our numerical results indicate that cVQE with the circuit ansatz of Fig.~\ref{fig:circuit_structure}(b) can indeed accurately prepare the lowest two eigenstates, and therefore allows for extracting the energy gap with high precision, even for systems with qubit numbers of $\mathcal{O}(100)$.

\section{The lowest eight eigenstates with three ancillary qubits\label{sec:Three-ancilary}}
So far, we demonstrated that cVQE can capture the first excited state with high precision, which allows for calculating the vector mass gap of the lattice Schwinger model. To obtain the scalar mass, one has to compute the next zero-momentum excitation in the spectrum, which requires determining more excited states. In this section, we demonstrate that the cVQE approach allows for reliably preparing multiple excited states and for identifying the dispersion relation of the lattice model, which in turn makes it possible to identify the scalar state.

\subsection{Energy momentum dispersion for small size systems}
Let us add two more ancillary qubits into the system, so that we can obtain the eight lowest eigenstates in the ideal case. We start with eight physical qubits and three ancillary qubits. We utilize the brick-wall circuit ansatz with universal SO(4) gates shown in Fig.~\ref{fig:circuit_structure}(a). In our simulation, we observe a drastic drop in infidelity when the number of ansatz layers reaches $15$. Therefore, we present the results for $16$ layers in this subsection.

\begin{figure}[!htp]
    \centering
    \includegraphics[width=0.4\textwidth]{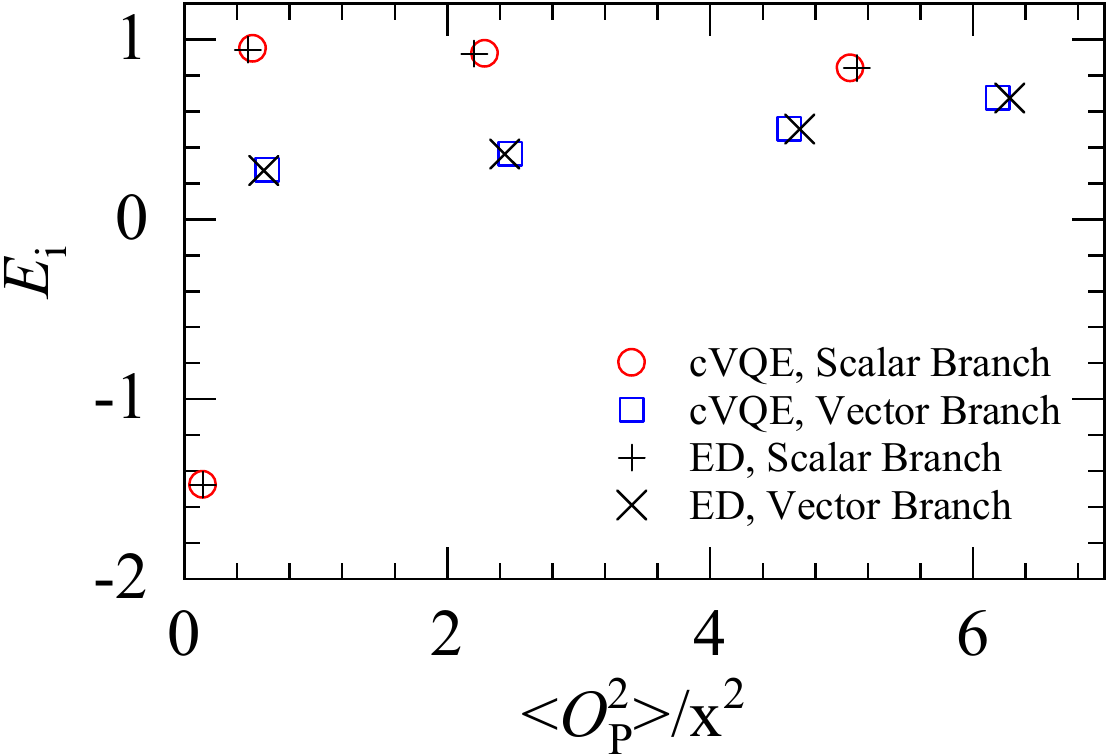}
    \caption{
     The energy pseudomomentum dispersion for eight physical qubits when $x=0.64$, $m_{\mathrm{lat}}/g=0.125$ and $l=0$ with 16 gate layers. The blue squares (red circles) represent the vector (scalar) branch from cVQE. The black "$\cross$" ("+") represent the vector (scalar) branch from ED for comparison. The different branches are determined by the phase of spin transformation $\langle S_R \rangle$, as introduced in subsection \ref{subsec:quantitiesdef}.
     }
    \label{fig:l0m0.125energymomentum}
\end{figure}

Figure~\ref{fig:l0m0.125energymomentum} shows the energy pseudomomentum dispersion for eight physical qubits obtained for $x=0.64$, $m_{\mathrm{lat}}/g=0.125$ and zero background electric field. As discussed in Sec.~\ref{subsec:quantitiesdef}, the scalar and vector branches can be identified by the phase $\phi(\langle S_R \rangle)$. The blue squares (red circles) represent the vector (scalar) branch from cVQE, where $\phi\approx\pi$ ($\phi\approx 0$). For comparison, we also present the ED results in the same figure, where the black `$\cross$' (`+') represents the vector (scalar) branch. Comparing the data from cVQE and ED we observe very good agreement for energies, pseudomomentum and $\phi$. The fidelity of all states is close to $1$. The data demonstrate that the cVQE is suitable for determining a larger number of excited states. The results are similar for the $l=0.125$ case, and we omit to display them here.

Compared to previous numerical simulations of the model, we observe that the vector branch behaviour is similar to that in Ref.~\cite{banuls2013mass} obtained with high-precision MPS simulations. However, the first scalar candidate does not have a pseudomomentum close to zero. As we will see in the next subsection, this discrepancy likely emanates from the larger lattice spacing used in this simulation.

\subsection{Increasing the system size}
The next step is to scale up the system size while maintaining the number of ancillary qubits unchanged. Here, we focus on sixteen physical qubits with three ancillary qubits. When utilizing the ansatz from Fig.~\ref{fig:circuit_structure}(a) and Fig.~\ref{fig:circuit_structure}(b), we found that some low-energy eigenstates were skipped and instead we obtained some of the next higher excited states. In our simulation, we found that this phenomenon could not be avoided, even with the number of layers in the ansatz reaching the order of the system size. The problem that some of the eigenstates are skipped was also observed in Ref.~\cite{xu2023concurrent}, where adding one more ancillary qubit, i.e., four ancillary qubits, was found to address the problem. Here, we implement a different solution by introducing universal multi-qubit gates acting on more than two qubits that ensure enough expressibility of the ansatz to capture all low-energy states we are interested in. The trade-off is that with this approach a larger number of parameters have to be optimized. In the following simulations with three-ancillary qubits, we adopt the simplest case of the universal three-qubit gates to balance expressivity and the number of parameters to be optimized.

\begin{figure}[htp!]
    \centering
    \includegraphics[width=0.4\textwidth]{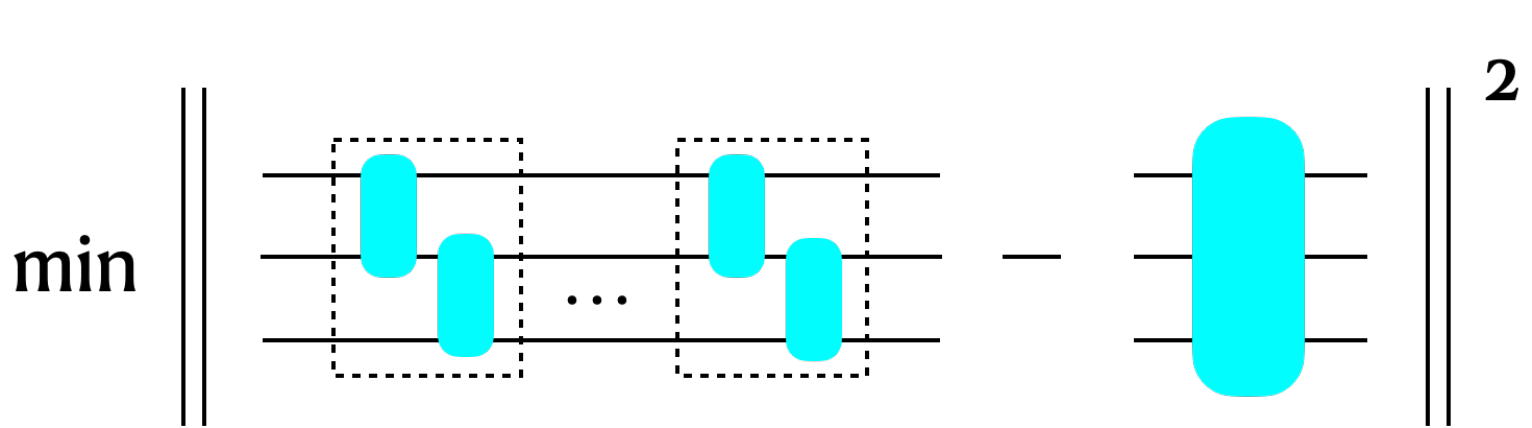}
    \caption{
     The decomposition of a universal SO(8) three-qubit gate. The gate is decomposed by minimizing the norm distance between the sequential application of universal SO(4) gates and the original three-qubit gate. Each dashed box represents a single layer of gates within the sequence, with the number of layers being tuned based on the error of the optimization.}
     \label{fig:gate_decomposition_so8}
\end{figure}

Figure~\ref{fig:circuit_structure}(c) introduces the universal three-qubit gate ansatz with a ladder structure. In our numerical simulations, we follow a two-step process for the optimization of the parameters. The first step is to construct the universal SO(8) gates according to Eq.~\eqref{Eq:so(n)exp}. This construction allow us to optimize the parameters via optimizing the strictly triangular $A$ matrix. We utilize a similar optimization strategy as in Sec.~\ref{subsec:cVQElarge} to simulate large systems with three-ancillary qubits. At the beginning we impose translational symmetry for 1000 iterations with $\lambda=1$ per site. Thereafter, the symmetry constraint is released and 2000 iterations are performed to optimize all gates independently with half the value for $\lambda$. The second step is to decompose the optimized multi-qubit gates by minimizing the square of the Frobenius norm between the targeted multi-qubit gate and the universal two-qubit gate ansatz. As shown in Fig.~\ref{fig:gate_decomposition_so8}, we utilize the universal SO(4) gates with ladder structure to decompose the targeted three-qubit gates. 
The universal two-qubit gates inside the dashed black box are grouped as one gate layer. The number of layers can be tuned based on the desired threshold of the difference in the Frobenius norm. In the following simulations, the number of two-qubit gate layers are chosen large enough to ensure a final Frobenius norm below $10^{-10}$.

\begin{figure}[htp!]
    \centering
    \includegraphics[width=0.4\textwidth]{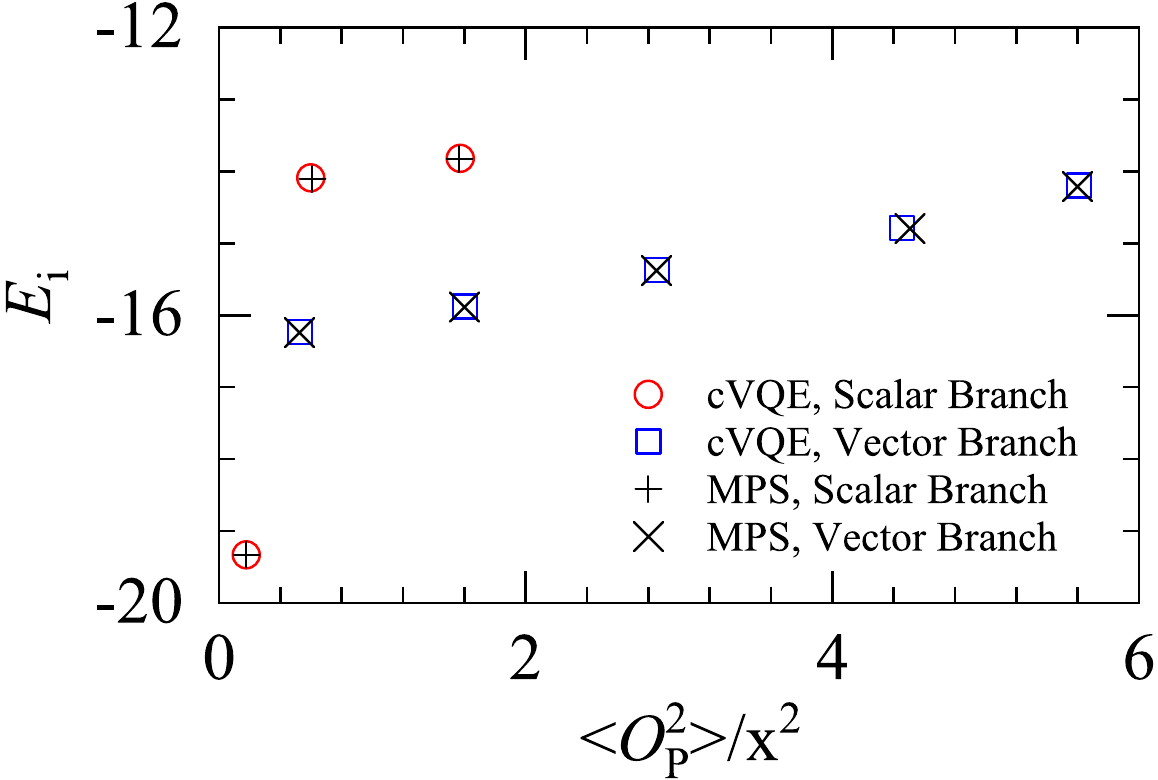}
    \caption{
     The energy pseudomomentum dispersion for 16 physical qubits with $x=2.56$, $m/g=0.125$, $l=0$ and 8 layers. The blue squares (red circles) represent the vector (scalar) branch from cVQE. The black "$\cross$" ("+") represent the vector (scalar) branch from MPS (bond dimension 40) for comparison. The different branches are identified by the phase of spin transformation $\langle S_R \rangle$, as introduced in subsection~\ref{subsec:quantitiesdef}.
     }
     \label{fig:N16l0m0.125energymomentum}
\end{figure}

First, let us focus on case of vanishing background electric field. We again focus on a dimensionless lattice volume of $10$, which corresponds $x=2.56$ for the chosen system size, and  $m_{\mathrm{lat}}/g=0.125$. The energy-pseudomomentum dispersion is obtained from the ansatz in Fig.~\ref{fig:circuit_structure}(c) where we find 8 layers are enough to determine the lowest eight eigenstates to fidelity higher than $0.99$. When decomposing the universal three-qubit gates, we empirically find that four universal two-qubit-gate layers are enough. Figure~\ref{fig:N16l0m0.125energymomentum} shows the energy-momentum dispersion results using the above conditions. 
Just as before, by examining the phase of $\langle S_R \rangle$, the eight eigenstates are divided into vector and scalar branches, whose symbols are blue squares and red circles, respectively. As the data reveals, for sixteen physical qubits, we observe a similar picture for the ground state and the vector branch as in the previous subsection. Above the ground state, which has approximately vanishing pseudomomentum, we find the scalar state and subsequently its momentum excitations. Comparing the scalar branch from Fig.~\ref{fig:N16l0m0.125energymomentum} to the one obtained for smaller system sizes (Fig.~\ref{fig:l0m0.125energymomentum}), we observe that the scalar branch now behaves similar to the vector branch: the first scalar state above the ground state has a pseudomomentum close to zero, and with increasing energy, the pseudomomentum is growing. This is the expected behaviour observed in previous numerical simulations~\cite{banuls2013mass}. Since we keep a constant dimensionless lattice volume $N/\sqrt{x}$ of $10$ in our simulations, going to a larger system size of 16 qubits corresponds to decreasing the lattice spacing. Our data indicate that this behaviour of the vector branch observed in Fig.~\ref{fig:l0m0.125energymomentum} is caused by finite lattice effects. 

Comparing the cVQE results in Fig.~\ref{fig:N16l0m0.125energymomentum} to MPS data with bond dimension 40, where the vector (scalar) branch is denoted as black `$\cross$' (`+'), we observe excellent agreement. To corroborate the agreement observed for the energy, we present the fidelity between the eigenstates obtained from the cVQE and the eigenstates from the direct MPS calculation in the second column of Table~\ref{tab:fidelity}. All values of the fidelity are higher than $0.99$. This agreement indicates that cVQE with the universal three-qubit gates can capture the lowest eight states and give the correct energy-momentum dispersion to high precision for the vanishing background electric field.

\begin{figure}[htp!]
    \centering
    \includegraphics[width=0.4\textwidth]{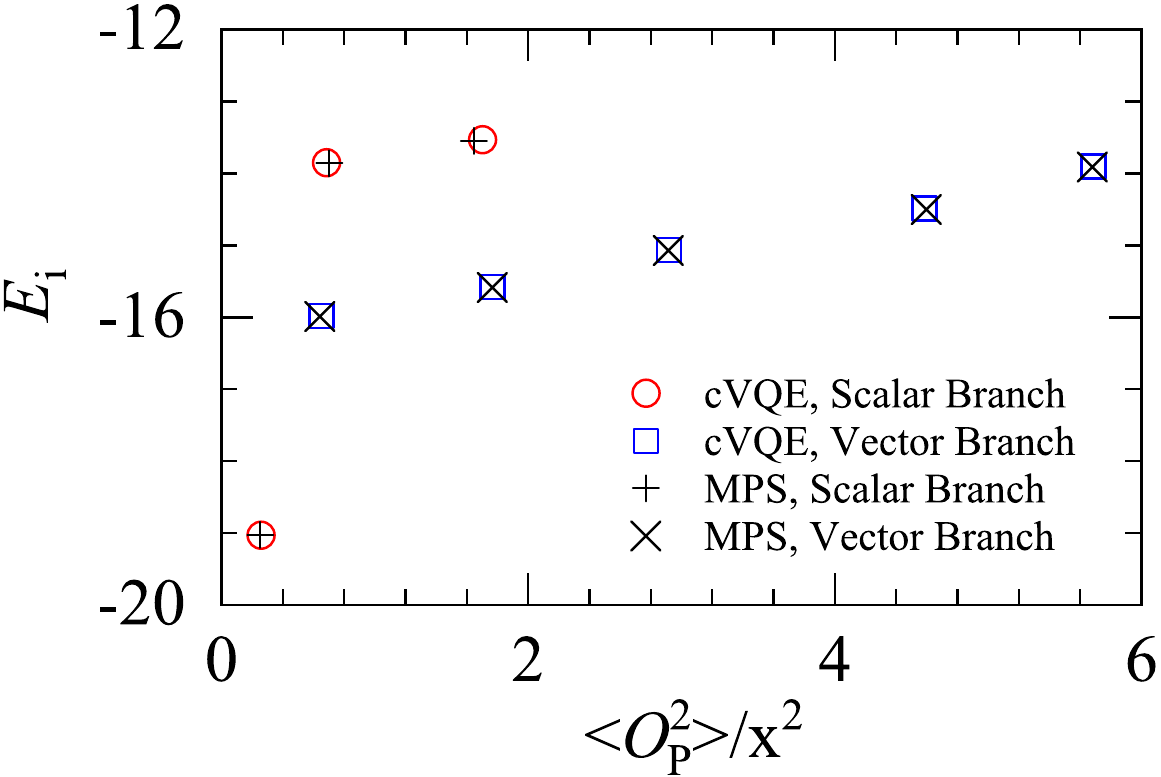}
    \caption{
        The energy pseudomomentum dispersion for 16 physical qubits with $x=2.56$, $m/g=0.125$, $l=0.125$ and 8 layers. The blue squares (red circles) represent the vector (scalar) branch from cVQE. The black "$\cross$" ("+") represent the vector (scalar) branch from MPS (bond dimension 40) for comparison. The different branches are identified by the phase of spin transformation $\langle S_R \rangle$, as introduced in subsection~\ref{subsec:quantitiesdef}.
     }
     \label{fig:N16l0.125m0.125energymomentum}
\end{figure}

To benchmark the cVQE in the presence of a nonvanishing topological term, a regime where conventional MC approaches suffer from the sign problem, we investigate $l = 0.125$. We choose the same parameters for the lattice mass and $x$. Again, we apply 8 layers of universal three-qubit gate layers for the circuits. The three qubit gates are then decomposed into universal two-qubit gates as before using up to four layers of universal two-qubit gates. The results for the energy-pseudomomentum dispersion are shown in Fig.~\ref{fig:N16l0.125m0.125energymomentum}, where we use the same marker convention as in the case of $l=0$. 
Also for the case of nonvanishing background electric field, we observe a vector branch above the ground state before finding the scalar branch of excitations. The data for the dispersion from the cVQE are in very good agreement with the numerical results from a direct MPS calculation, showing that cVQE allows for obtaining excited states with good precision in a regime where conventional MC methods suffer from the sign problem. This is further confirmed by looking at the fidelity between the states obtained from the MPS calculations and the cVQE, shown in the third column of Table~\ref{tab:fidelity}. Again, we find the fidelity of all states are higher than $0.99$, indicating that cVQE with the universal three-qubit gate ansatz from Fig.~\ref{fig:circuit_structure}(c) can precisely simulate the lowest eight eigenstates in the presence of a topological term for large systems.

\renewcommand\arraystretch{1.4}
\begin{table}
    \begin{center}
        \caption{The fidelity of the lowest eight eigenstates for a system with 16 sites, $x=2.56$ and $m_{\mathrm{lat}}/g=0.125$, for vanishing and non-vanishing background electronic field. These results are obtained with 8 ansatz layers using the universal three-qubit gates.}
        \setlength{\tabcolsep}{1.55mm}
        \begin{tabular}{p{1.2cm}<{\centering}|p{2.5cm}<{\centering}|p{2.5cm}<{\centering}}
        \hline
        \hline
        State & $F_i$ for $l$=0 & $F_i$ for $l$=0.125\\
        \hline
        GS & 0.9999090203 & 0.9998516139\\
        \hline
        $1st$ ES & 0.9994775108 & 0.9992610969\\
        \hline
        $2nd$ ES & 0.9994892349 & 0.9992590285\\
        \hline
        $3rd$ ES & 0.9993555690 & 0.9992846633\\
        \hline
        $4th$ ES & 0.9981458180 & 0.9981867413\\
        \hline
        $5th$ ES & 0.9976975423 & 0.9970726774\\
        \hline
        $6th$ ES & 0.9970323007 & 0.9953149883\\
        \hline
        $7th$ ES & 0.9976732373 & 0.9959708763\\
        \hline
        \hline
        \end{tabular}
        \label{tab:fidelity}
    \end{center}
\end{table}

\section{Determining the Mass Shift via the Energy Gap\label{sec:massshift}}
As discussed in Sec.~\ref{subsec:quantitiesdef}, the lattice discretization leads to an additive mass renormalization $m_s(x,N,l)$. Reference~\cite{dempsey2022discrete} analytically derived the mass shift for the staggered discretization of the Schwinger model with periodic boundary conditions, showing that $m_s(x,N,l)=g^2a/8$. Moreover, it was shown that taking the additive mass renormalization into account greatly improves the convergence towards the continuum limit. The work has been extended in Refs.~\cite{Angelides2022,angelides2023computing}, where the authors developed a strategy for determining the additive mass renormalization for $l\neq 0$ based on the fact that the electric field density in the Schwinger model is expected to vanish at zero renormalized mass. However, the technique in Refs.~\cite{Angelides2022,angelides2023computing} is not directly applicable for $l=0$. Here, we present an alternative method based on the energy gap of the theory that works for any background electric field including $l=0$. As we demonstrated, cVQE allows for obtaining excited states with high precision, hence enabling us to determine the mass shift using the energy gap. 

\subsection{General approaches to extract the mass shift} 
In order to be able to follow a line of constant physics, e.g., for taking the continuum limit of an observable, it is necessary to account for the mass shift according to Eq.~\eqref{eq:mass_shift}. From that relation, one sees that for $m/g=0$ one can obtain the mass shift via
\begin{equation}
    \frac{m_s(x,N,l)}{g} = -\frac{m_{\mathrm{lat}}}{g} \bigg|_{m/g=0}.
    \label{Eq:MsMlat}
\end{equation}
For the continuum Schwinger model, some solutions are exactly known for vanishing bare fermion mass. For example, the electric field density is exactly zero for $m/g = 0$, because of the screening effect~\cite{coleman1976more,adam1996schwinger}. Reference~\cite{angelides2023computing} used this fact to determine the mass shift by finding the $m_{\rm{lat}}/g$ for which $F^{av}=0$ when $l \neq 0$. We refer to this method as the `Electric Field Density (EFD) method' in the following discussion.

Similarly, one can use the energy gap in the massless limit, which is exactly the mass of a free Schwinger boson shown in Eq.~\eqref{Eq:MassSchwingerBoson}. Choosing the lattice mass $m_{\rm{lat}}/g$ such that the mass gap of the lattice model $\Delta (m_{\rm{lat}}) = E_1 - E_0$ is equal to the Schwinger boson mass,
\begin{equation}
    \frac{\Delta (m_{\rm{lat}}/g)}{2 \sqrt{x}} = \frac{1}{\sqrt{\pi}},
    \label{Eq:MassGapBoson}
\end{equation}
one can identify the point of vanishing renormalized mass and determine the mass shift according to Eq.~\eqref{Eq:MsMlat}. Notice that in the expression above we rescaled the mass gap of the lattice model by $2\sqrt{x}$, which accounts for the factor $ag^2/2$ in the dimensionless Hamiltonian in Eq.~\eqref{Eq:SpinHamiltonian}. In the following discussion, we will refer to this approach as the `Gap method'. According to the results from mass perturbation theory in Eq.~\eqref{Eq:MassSchwingerPerturbation}, one expects a monotonic dependence for the energy gap as a function of the renormalized mass $m/g$ around zero. This observation allows for the bisection method to be used effectively to solve Eq.~\eqref{Eq:MassGapBoson}. 
The precision of the energy gap determines the resolution of the gap method. Compared to the EFD method, we expect it to be stable against the local noise in real quantum device simulations, since the energy gap is a global physical quantity.  

\subsection{Identifying the mass shift from the energy gap\label{subsec:gap_mass_shift}}
In this subsection we use the Gap method to determine the mass shift of different $x$. We first focus on validating the correctness of the gap method, thus we utilize direct MPS simulations instead of cVQE to compute the mass shift as this eliminates possible bias effects due to the ansatz choice. In MPS simulation, the bond dimension is set to keep the truncation error lower than $10^{-10}$. When computing the excited states with MPS, we first determine the ground state and then add a penalty term to shift it to an energy higher than the first excited state. After that, the calculation of the first excited state is identical to computing the ground state of the modified Hamiltonian, shown in Eq.~\eqref{eq:mpsgap}. The coefficient in the penalty term is chosen carefully to ensure the overlap between these two states is in the same order as the truncation error. The bisection process of $m_{s}/g$ starts from interval $[-0.16, 0]$ and terminates when the interval length is less than $10^{-8}$. After the bisection process, we regard the interval's middle value as the resulting $-m_{s}/g$.

\begin{figure}[!tp]
    \centering
    \includegraphics[width=0.4\textwidth]{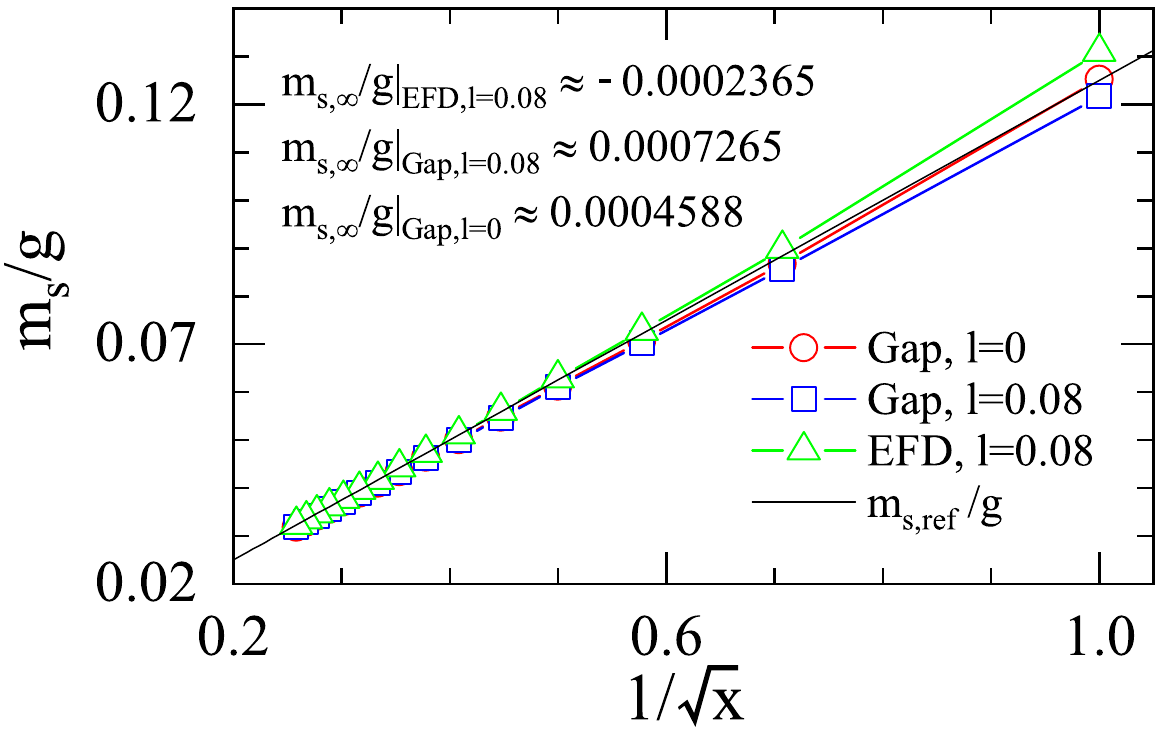}
    \caption{
    The behaviour of the mass shift $m_s/g$ as a function of $1/\sqrt{x}$ for $l=0$ (red circles) and $l=0.08$ (blue squares). The mass shift is estimated with the Gap method by solving Eq.~\eqref{Eq:MassGapBoson} for various $N$ using MPS and subsequently extrapolating to infinite volume for a fixed $x$. For comparison, the results from the EFD method are also presented for $l=0.08$ (green triangles). The black line represents the analytically computed value $1/8\sqrt{x}$ from Ref.~\cite{dempsey2022discrete} for periodic boundary conditions. 
    }
    \label{fig:MPSMassShift}
\end{figure}

The mass shift of different $x$ is determined to analyze the performance of the Gap method at different lattice spacings. For each $x$, we calculate the $m_{s}/g$ for a set of lattice sizes $N \in [100:20:400]$, and extrapolate in $\sqrt{x}/N$ i.e. inverse volume, by cubic polynomial fitting to obtain the mass shift at infinite volume. The fitted mass shifts of various $x \in [1,15]$ from the Gap method are shown in Fig.~\ref{fig:MPSMassShift}, including results for $l=0$ (red circles) and $l=0.08$ (blue squares). Using a linear fit for the data of $x \in [5:1:15]$, we extrapolate to $m_{s,\infty}/g$ for the infinite $x$ value. 
To have a well defined reference, we also present the infinite-volume mass shift determined by the EFD (green triangles) for the same $x$. The central four gauge links are chosen to compute the EFD to reduce the boundary effects. Again, we use the bisection method to determine the mass shift in the EFD method. The bisection strategy for each $x$ is the same as in the Gap method. Since the mass shift obtained from the bisection method remains unchanged when the system size $N$ exceeds $260$ for all $x$, we do not extrapolate and just pick the value at $N=400$ as the infinite-volume mass shift. As before, using a linear fit for the results of $x \in [5:1:15]$, we obtain the $m_{s,\infty}/g$ for infinite $x$.

The first observation in Fig.~\ref{fig:MPSMassShift} is that independently of the method, the mass shift approaches zero as expected when the continuum limit is approached and additive mass renormalization coming from the discretization of the theory onto the lattice is removed. Second, for small $x$ or equivalently large lattice spacings, we observe a difference between the analytical solution derived for periodic boundary conditions and our numerical data for systems with open boundary conditions having nonvanishing background electric field. This difference vanishes as $x$ increases and one approaches the continuum limit. Compared with the mass shift from the EFD method and analytical result, we conclude that the Gap method produces consistent results for large enough $x$. This behaviour demonstrates the feasibility of using the Gap method to extract the mass shift induced by the lattice effects.

\begin{figure}[!tp]
    \centering
    \includegraphics[width=0.4\textwidth]{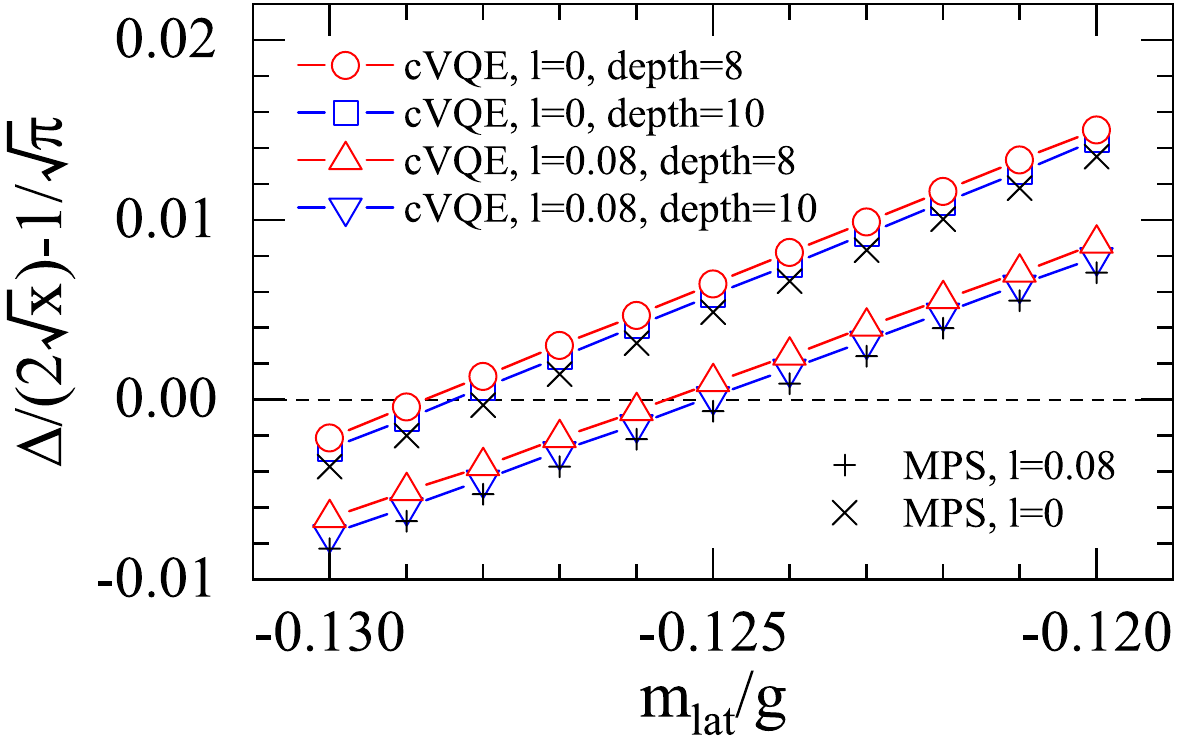}
    \caption{Results for $\Delta/(2\sqrt{x})-1/\sqrt{\pi}$ with $m_{\rm{lat}}/g$ for $40$ physical qubits and $x=1$ from cVQE (open symbols) and a direct MPS calculation ($\cross$, +) for comparison. The red circles (blue squares) and red triangular (blue lower triangle) markers correspond to results of $l=0$ and $l=0.08$ for eight (ten) layers of the ladder ansatz from Fig.~\ref{fig:circuit_structure}(b). The black "$\cross$" ("+") represents data from a direct MPS calculation with bond dimension $D=40$, serving as the reference value for $l=0$ ($l=0.08$). The dashed line marks $\Delta/(2\sqrt{x})-1/\sqrt{\pi} = 0$, its intersection with cVQE data provides an estimate of the mass shift up to a minus sign.
    }
     \label{fig:GapMass}
\end{figure}

While our direct MPS data demonstrate that the Gap method allows for determining the mass shift, a crucial question is whether the cVQE can yield precise enough data to determine the mass shift from the energy gap. For cVQE, the energy gap can have systematic errors related to both circuit structure and optimization strategy. In the following we will provide a proof-of-principle that the cVQE approach is suitable to determine the mass shift. To this end, we examine the energy gap for a series of $m_{\rm{lat}}/g$ values with typical $x$ and $l$. To extract the mass shift directly, we present the behaviour of $\Delta/(2\sqrt{x})-1/\sqrt{\pi}$ instead of the energy gap $\Delta$ with $m_{\rm{lat}}/g$. Figure~\ref{fig:GapMass} shows the results from cVQE for $l=0$ and $l=0.08$ with $N=40$. 
As an example, we choose $x=1$ giving a volume of $40$. The reason is that lower excited states are hard to distinguish for small $x$, making this a more rigorous validity check of cVQE's ability to extract the mass shift. 
It is worth mentioning the optimization scheme in more detail. For the data using 8 layers of the ansatz circuit, we first optimize the case with $m_{\rm{lat}}/g=0$ and $l=0$, following the strategy introduced in subsection~\ref{subsec:cVQElarge}. Then, the parameter values of the best-performing gates are used as initial values with an added small random for other values of $m_{\rm{lat}}/g$ and $l$. The best performance is identified as the lowest energy from the $11$ random initial parameters. For the case of $10$ layers ansatz, we take the initial parameters to be the optimal parameters from the ansatz with $8$ layers and again add a small random noise. Meanwhile, the parameters in the two new layers are initialized with just small random noise.

Looking at Fig.~\ref{fig:GapMass}, we can see that the values obtained for $\Delta/(2\sqrt{x})-1/\sqrt{\pi}$ for $10$ layers are closer to the MPS results than those for $8$ layers, indicating a more precise mass shift estimation in the former case. Combined with the intersection with $\Delta/(2\sqrt{x})-1/\sqrt{\pi} = 0$ in Sec.~\ref{subsec:gap_mass_shift}, we conclude that the precision is around $10^{-3}$ for $8$ or $10$ layers, including vanishing and non-vanishing background electric field. More ansatz layers are needed for a more precise estimation of the mass shift since with more layers the results are close to the MPS reference results. Another interesting observation is that $\Delta/(2\sqrt{x})-1/\sqrt{\pi}$ changes linearly with $m_{\rm{lat}}/g$. This linear relation can be easily understood if we consider the expansion of Eq.~\eqref{Eq:MassSchwingerPerturbation} around the $m_{r}/g = 0$. For small $m/g$, the $M_s/g$ is linearly dependent on $m/g$ in the continuum and therefore on $m_{\rm{lat}}/g$ around $-m_s(x,N,l)/g$ on the lattice. We conclude that cVQE can simulate the energy gap $\Delta$ precisely enough and, therefore, can be used to extract the mass shift.

\section{Demonstration on a quantum device\label{sec:experiment}}
In the previous section, we have demonstrated the feasibility of simulating excited states of the Schwinger model using cVQE. However, it remains unclear whether the near-term devices could perform cVQE to prepare excited states of the Schwinger model efficiently. The answer to this question is closely related to the ability to prepare the excited states on near-term devices. This motivates us to prepare and measure the quantum states directly on a near-term device. This section reports the results on IBM's quantum device $ibm\_algiers$.   

\subsection{Strategy to prepare excited states on near-term quantum devices}
This subsection starts by introducing the strategy to prepare states on real devices. The error of near-term devices limits the possible circuit depth. Therefore, we need to balance the circuit expressibility and the accessibility of low-error results. Moreover, the required circuit depth grows with system size, as our previous simulations show. Consequently, we proceed with a preliminary resource estimation by conducting a classical simulation of cVQE. To maintain the complexity of the states, we fix $m_{\rm{lat}}/g=0.333$ and $l=0.5$, which is the critical point in the thermodynamic limit.
In the simulations, we found that the brick-wall circuit ansatz with two layers, i.e., eight CNOT layers, can produce states with fidelity of more than $0.9999$ when $N=4$ and one ancillary qubit is used. The low CNOT depth and the sufficient fidelity make this case an ideal example to test the ability to prepare excited states of the Schwinger model on a near-term device. Therefore, we adopt this brick-wall ansatz with four physical qubits and a single ancillary qubit in our subsequent implementation.

\begin{figure*}[!htbp]
    \centering
    \includegraphics[width=0.8\textwidth]{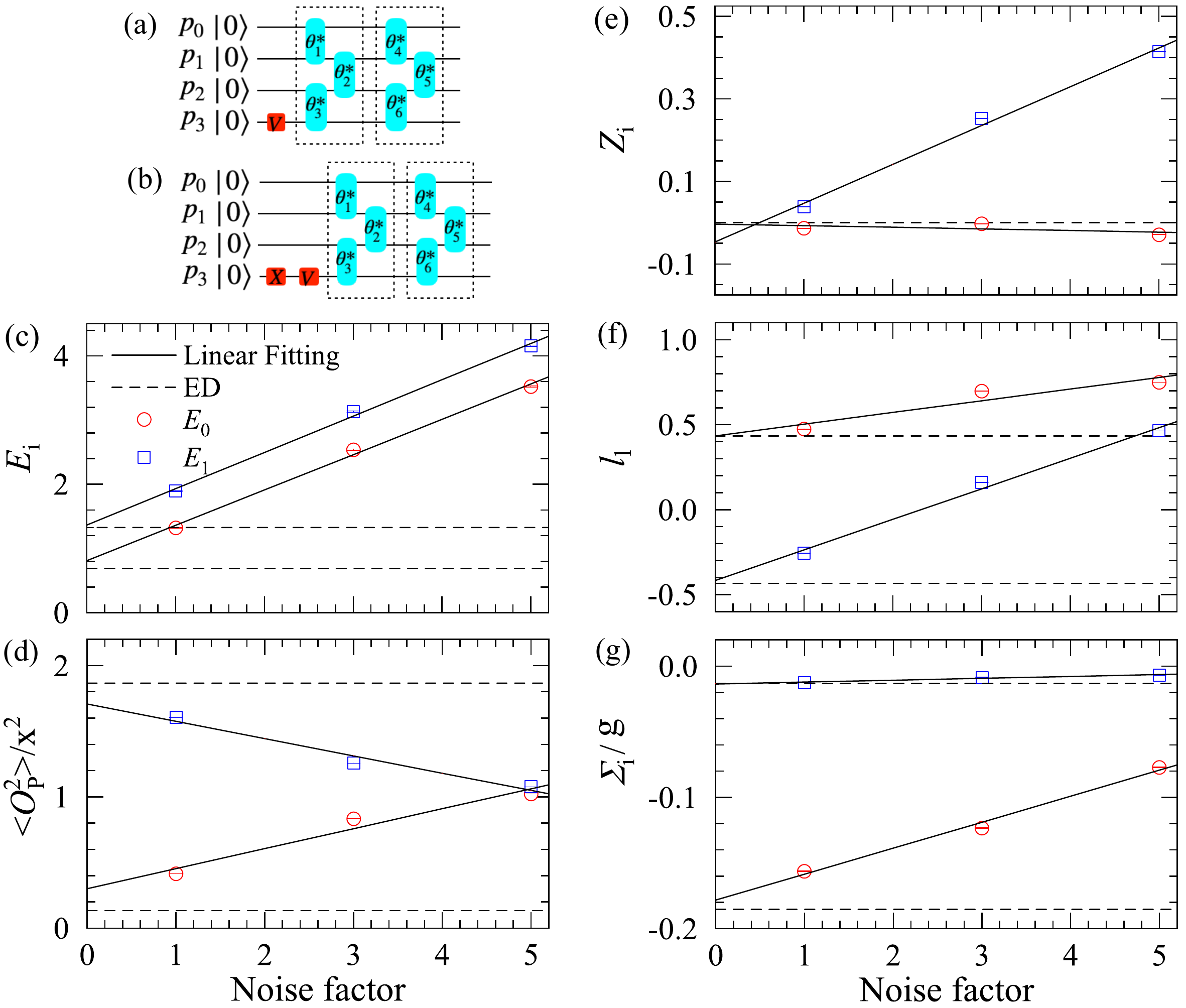}
    \caption{Circuit setup and results of the ground state (red circles) and the first excited state (blue squares) on the IBM device $ibm\_algiers$. The $N=4$ Schwinger model with $x=0.16$, $m_{\rm{lat}}/g=0.333$ and $l=0.5$ is used here as a demonstrative case. (a) and (b) provide the alternative equivalent circuit realization instead of the circuit in cVQE to prepare the ground state and the first excited state. The rotation gate $V$ is determined by the final diagonalization step in cVQE. We show the measurement results with error bars from $100000$ shots. The measured quantities include (c) energy $E_i$, (d) square of pseudomomentum $O_P/x$, (e) total charge $Z_i$, (f) expectation value of central link $l_1$, and (g) chiral condensate $\Sigma_i/g$. The solid black line give the fitting results from noise level 1, 3, and 5 with linear regression. Meanwhile, the black dashed lines mark ED results for comparison.}
    \label{fig:DeviceResults}
\end{figure*}

Instead of running the circuit many times, we choose to implement inference runs to demonstrate the feasibility of cVQE in obtaining excited states of Schwinger model on near-term devices. When operating the inference runs, we follow a two-step strategy, namely the classical determination of optimized gate parameters and the quantum preparation of states. First, the classical optimization is accomplished by the main procedures in Sec.~\ref{sec:cvqe}. Then, we separately prepare $2^{N_a}$ eigenstates using the optimized gates from given initial states in order to decrease the CNOT gate layers. Figure~\ref{fig:DeviceResults}(a) and (b) show the detailed circuits to prepare the ground state (GS) and the first excited state (1st ES) in our inference runs. Here, $\theta_i^*$ are the optimal gate parameters determined by the classical cVQE simulation. While gate $V$ is the additional rotation to obtain the eigenstates. The two initial states are identified as $|0000\rangle$ and $|0001\rangle$, from the fact that the initial state used in cVQE is $\frac{1}{\sqrt{2}}(|00000\rangle + |00011\rangle)$. In addition, we demonstrate another approach to the inference runs including ancillary qubits and put the results in the Appendix~\ref{sec:prepare_states_ancillary}. 

The zero-noise extrapolation (ZNE) is adopted to mitigate the effects of the device noise~\cite{giurgica2020digital}. For the whole circuit gate $U$, ZNE includes several extra $U^{\dagger}U$ pairs on the same device to increase incoherent noise while keeping the entire gate unchanged. This folding process creates a series of equivalent gates, i.e., $U, UU^{\dagger}U, UU^{\dagger}UU^{\dagger}U,...$, which are marked as different noise levels $1,3,5,...$. For circuits with different noise levels, measurements in the computational basis are carried to estimate physical quantities such as energy. Then the theoretical zero-noise quantities can be extrapolated to noise level $0$ with a fitting function. In our experiments, we utilize linear fitting to extract the extrapolated zero-noise quantities from measurements in noise levels 1, 3, and 5.

\subsection{Demonstration Results}
In this subsection, we report the experimental results of several quantities. For every physical quantity with various noise levels, we use $100000$ measurements to obtain an estimation of the expectation value. These results are shown in Fig.~\ref{fig:DeviceResults}(c-g). In each figure, we mark the results of the ground state (first excited state) with the red circles (blue squares). Further, the black solid and dashed lines are introduced to visualize the results of linear fitting and ED, respectively. To compare the observations from ZNE and ED directly, Table~\ref{tab:devicedata} is assigned to show the extrapolated zero-noise observable values and their reference ED data.

Figure~\ref{fig:DeviceResults}(c) shows the energy behaviour of the ground state and the first excited state. The energy from ZNE agrees with ED and we have sufficient accuracy to distinguish the ground state from the first excited state. This agreement is also observed in Fig.~\ref{fig:DeviceResults}(d), which describes the pseudomomentum. The zero total charge condition is tested by measuring the overall spin along the $z$-axis. The small value in Fig.~\ref{fig:DeviceResults}(e) indicates that the noise breaks the condition of vanishing total charge but to a limited extent. In addition, two other representative observables, the electric field on the central link and the chiral condensate, are measured to illustrate the accessibility of general physical quantities on the device. As we can see in Fig.~\ref{fig:DeviceResults}(f) and (g), both of them agree well with ED when considering ZNE.
These strong agreements between ZNE and ED results demonstrate the feasibility of preparing excited states on near-term devices. These promising results strongly indicate that current quantum devices have the potential to simulate the excited states of the Schwinger model.

\renewcommand\arraystretch{1.4}
\begin{table*}
    \begin{center}
        \caption{The extrapolated zero-noise results of the ground state (GS) and the first excited state (1st ES) of 4 sites system, with $x=0.16$, $m_{\rm{lat}}/g=0.333$, and $l=0.5$. These results are part of Figure~[\ref{fig:DeviceResults}].}
        \setlength{\tabcolsep}{1.55mm}
        \begin{tabular}{c|c|c|c|c}
        \hline
        \hline
        Quantities & GS ($ibm\_algiers$) & GS (ED) & $1st$ ES ($ibm\_algiers$) & $1st$ ES (ED) \\
        \hline
        Energy ($E_i$) & 0.8084440813 & 0.6872150210 & 1.3640642173 & 1.3253490258\\
        \hline
        Square of pseudo-momentum ($\langle O^2_P \rangle/x^2$) & 0.3006058333 & 0.1335730366 & 1.7067800000 & 1.8663884575\\
        \hline
        Total charge ($Z_i$) & -0.0033366667 & 0 & -0.0468950000 & 0 \\
        \hline
        Electric field on the central link ($l_1$) & 0.4343491667 & 0.4331668567 & -0.4165708333 & -0.4331160521\\
        \hline
        Chiral condensate ($\Sigma_i/g$) & -0.1782994167 & -0.1852093713 & -0.0136271667 & -0.0131974491\\
        \hline
        \hline
        \end{tabular}
    \label{tab:devicedata}
    \end{center}
\end{table*}

\section{Summary and Outlook} 
\label{sec:summary}
This paper introduced a novel scheme to simultaneously prepare many excited states of the Schwinger model. Using the cVQE, together with universal SO(4) and SO(8) gate circuit ansätze and the well-designed optimization strategies, up to the lowest eight eigenstates have been prepared with high fidelity, including both the case of vanishing and nonvanishing background electric field. By applying tensor-network techniques which compress the circuit states, we confirmed that this scheme can precisely prepare the first excited states of systems up to $\mathcal{O}(100)$ qubits. These high-fidelity results enable us to determine the physical observables, such as the energy gap and the energy-pseudomomentum dispersion, and prepare excited states on near-term quantum devices. As a demonstration, we prepared the ground and first excited state of the four sites Schwinger model on IBM's quantum device $ibm\_algiers$, and the measurement of various quantities have shown good agreement with ED.  

Considering the ability to calculate the energy gap precisely, we propose using the energy gap to determine the additive mass renormalization of the Schwinger model emanating from its lattice discretization. Combined with the MPS simulation, we confirmed the feasibility of this method in estimating the mass shift. This successful application enables us to access the mass shift directly on a quantum computer. 

Based on the findings of this paper, some extensions of our scheme remain to be explored in future. In principle, adding more ancillary qubits to simulate more excited states of the Schwinger model is possible. When performing cVQE on near-term quantum devices, the noise effect should be considered carefully. From the simulation perspective, accounting for noise effects necessitates simulating a density matrix instead of a pure state, which requires us to effectively integrate noise models with the cVQE scheme. Once the excited states of large systems are successfully prepared on a quantum device, the dynamics of these states can be explored~\cite{chai2312entanglement}. Therefore, the combination of state preparation and dynamics simulations would be another extension. We leave these topics for future study.

\acknowledgments
Yibin Guo is grateful to Yahui Chai for the helpful discussions. This work is funded by the European Union's Horizon Europe Framework Programme (HORIZON) under the ERA Chair scheme with grant agreement no.\ 101087126. This project has received funding from the European Union's Horizon 2020 research and innovation programme under the Marie Skłodowska-Curie grant agreement No 101034267.
This work is supported with funds from the Ministry of Science, Research and Culture of the State of Brandenburg within the Centre for Quantum Technology and Applications (CQTA). 
\begin{center}
    \includegraphics[width = 0.08\textwidth]{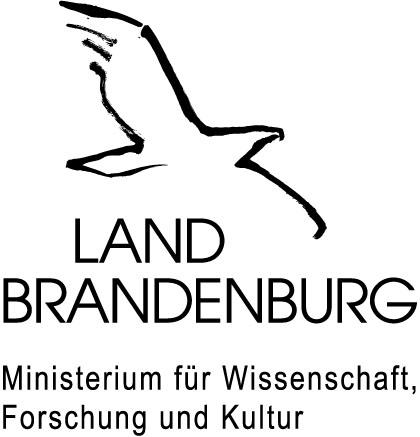}
\end{center}

\appendix

\section{The lowest four eigenstates with two ancillary qubits\label{sec:Two-ancilary}}
In this section, we report typical results of two ancillary qubits, including vanishing and non-vanishing background electric field. This section also discusses the energy and fidelity convergence of a small system, i.e., eight physical qubits.

\subsection{Vanishing Background Electric Field}
Two ancillary qubits allow for determining the lowest four eigenstates in the ideal case. As illustrative results, we only focus on the case with $volume = 10$, i.e., $x=0.64$ in this subsection. The maximum number of optimization steps is set to $4000$, enough to ensure the results converge to high precision. As shown in this section, we picked the lowest mean energy results from a group of eleven random initial parameters. Additionally, the $\lambda$ in penalty term~\eqref{Eq:PenaltyHamiltonian} is set to $8$ to ensure the condition $|\langle \sum_i \sigma_i^z \rangle| \approx 0$.

\begin{figure}[!htbp]
    \centering
    \includegraphics[width=0.45\textwidth]{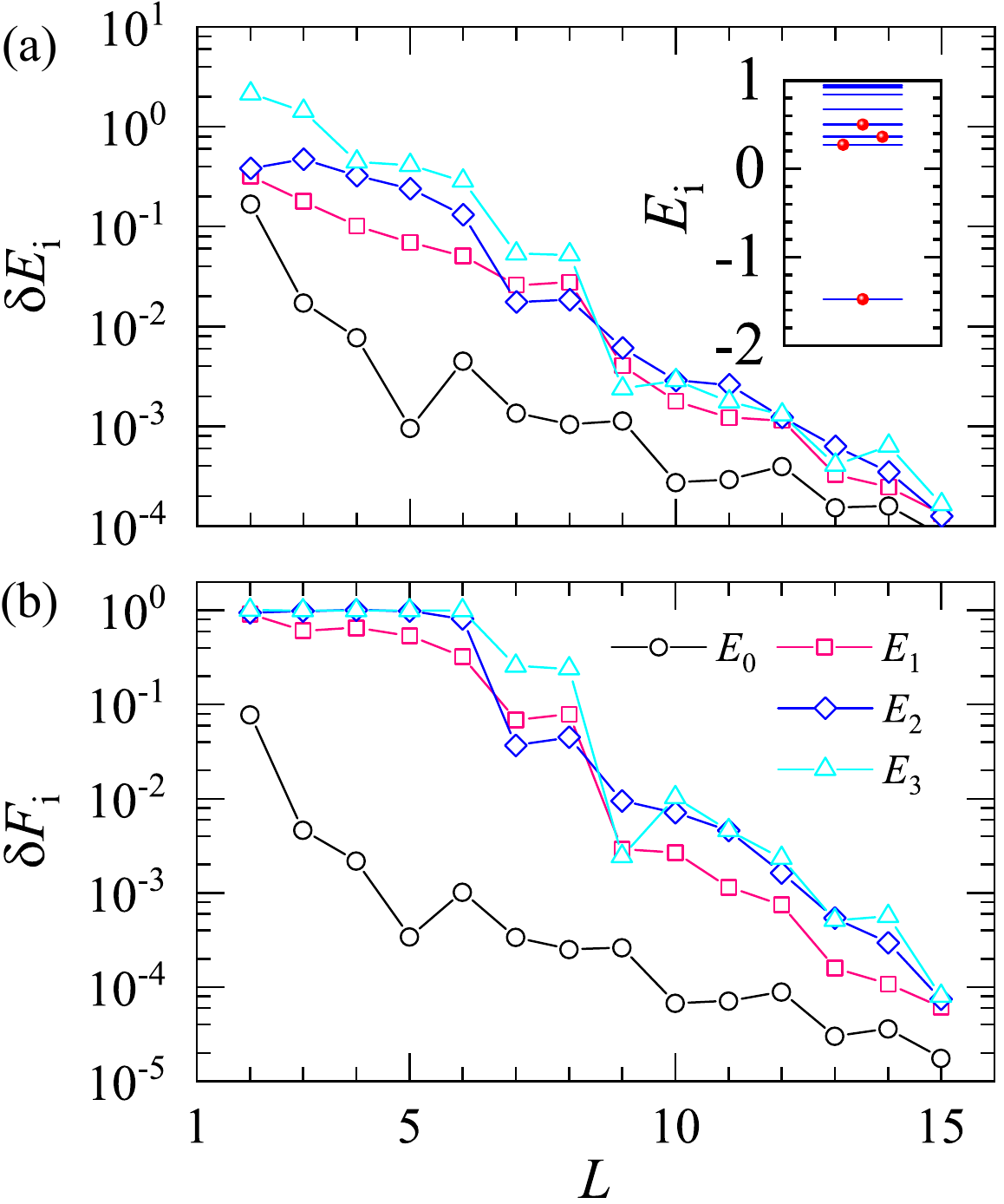}
    \caption{
     The convergence behaviour of (a) $\delta E_i = E_i - E_i^{ed}$ and (b) infidelity $\delta F_i$ with the number of layers $L$. The inset in (a) shows the energy levels, where the red dots represent energy from cVQE with fifteen gates layers and blue lines are the energy calculated by ED.
     There are eight physical and two ancillary qubits. We choose the parameters as: $volume = 10$ ($x=0.64$), $m_{\rm{lat}}/g=0.125$, with vanishing background electric field. The results are selected from eleven random seeds according to the lowest expectation value of Eq.~\eqref{Eq:SpinHamiltonian} with constraint Eq.~\eqref{Eq:PenaltyHamiltonian}. 
     }
     \label{fig:EFN8A2l0m0.125}
\end{figure}

The results are determined by both the circuit structure and the circuit depth. We utilize the circuit ansatz in Fig.~\ref{fig:circuit_structure}(a) to simulate the case of eight physical qubits and two-ancillary qubits. Figure~\ref{fig:EFN8A2l0m0.125}(a) shows how the energy difference $\delta E_i = E_i - E_i^{ed}$ of the four states converges with the circuit depth. The $\delta E_i$ decreases as the ansatz layers grow. The inset in Fig.~\ref{fig:EFN8A2l0m0.125}(a) shows the energy levels. The red dots represent the simulation results with fifteen layers, while the blue lines represent the ED reference results. This inset gives intuition about the accuracy and the energy level distribution. Notice that the first and the second excited state are very close, which requires a high resolution of the energy levels to distinguish them. With cVQE and circuit ansatz Fig.~\ref{fig:circuit_structure}(a), we can determine the energy levels to a precision $10^{-4}$ and therefore overcome this challenge successfully.

To further demonstrate the accuracy we obtained, we show the convergence behaviour of the infidelity $\delta F_i = 1 - F_i$ with gate depth in Fig.~\ref{fig:EFN8A2l0m0.125}(b). One interesting observation is that the infidelity of the three excited states drops dramatically when the number of layers reaches nine. This change in behaviour means the circuit has enough expressibility after that point. Besides, uniform convergence is guaranteed by the choice of the loss function when the circuit structure has enough expressibility. The behaviour confirms this prediction that all four states converge uniformly in both energy and infidelity. For fifteen layers, the infidelity reaches an order of $10^{-4}$ for all states, where the variance of the Hamiltonian for each state is less than $10^{-3}$. These two facts indicate that we indeed can simulate the lowest four eigenstates with cVQE and ansatz from Fig.~\ref{fig:circuit_structure}(a) with high precision.

\subsection{Nonvanishing Background Electric Field}
The performance of cVQE in the nonvanishing background electric field remains to be explored. In this subsection, we continue to discuss the two ancillary qubits results in non-vanishing background electric field $l = 0.125$. We consider eight physical qubits system with volume = 10, i.e., $x=0.64$. The penalty term is set as $8$ to constraint the condition $|\langle \sum_i \sigma_i^z \rangle| < 10^{-6}$. We take 4000 optimization steps to ensure convergence. The results are selected out of the eleven random seeds for each parameter and gate layer according to the lowest mean energy criterion.

\begin{figure}[!tb]
    \centering
    \includegraphics[width=0.45\textwidth]{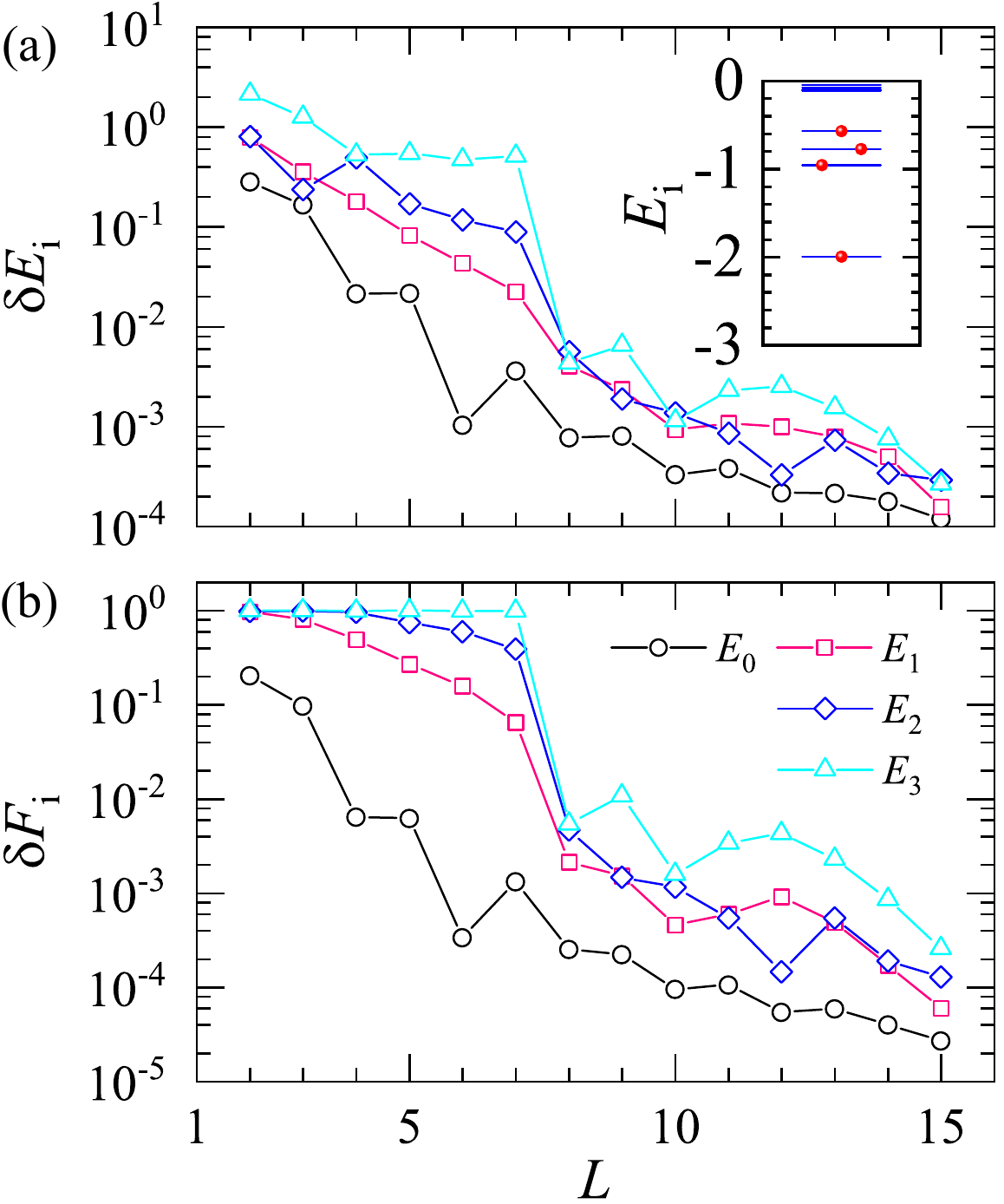}
    \caption{
     The convergence behaviour of (a) $\delta E_i = E_i - E_i^{ed}$ and (b) infidelity $\delta F_i$ with the number of layers $L$. The inset in (a) shows the energy levels, where the red dots represent energy from cVQE with fifteen gates layers and blue lines are the energy calculated by ED. The figures show the lowest four eigenstates for eight physical qubits and two ancillary qubits. Here we choose $x = 0.64$, $m/g = -0.125$, and $l=0.125$. The results are selected from eleven random seeds according to the lowest expectation value of Eq.~\eqref{Eq:SpinHamiltonian} with constraint Eq.~\eqref{Eq:PenaltyHamiltonian}.
     }
     \label{fig:EFN8A2l0.125m-0.125}
\end{figure}

First, we give the typical results of two ancillary qubits. As an example for negative mass, we take $m/g=-0.125$. Figure~\ref{fig:EFN8A2l0.125m-0.125}(a) shows the convergence behaviour of the energy difference $\delta E_i = E_i - E_{\mathrm{ed}}$ with gate depth $L$. As $L$ increases, the average $\delta E_i$ decreases for all four states. The $\delta E_i$ of all the four states reaches a precision of $10^{-4}$ for a depth of fifteen. The inset in Fig.~\ref{fig:EFN8A2l0.125m-0.125}(a) gives the energy levels of the fifteen layers. The red dots represent the simulation results, while the blue lines represent the ED results. Figure~\ref{fig:EFN8A2l0.125m-0.125}(b) shows the convergence behaviour of infidelity $\delta F_i = 1 - F_i$. 
Like the vanishing background electric field case, the infidelity $\delta F_i$ drops dramatically when gate layers reach eight. The dropping behaviour indicates the circuit has enough expressibility after eight layers. At fifteen gates layers, the infidelity $\delta F_i$ of all states is less than $10^{-4}$, where the variance of the Hamiltonian of the four states reaches an order of $10^{-3}$. The cVQE with circuit ansatz in Fig.~\ref{fig:circuit_structure}(a) can simulate the lowest four states to high precision for the nonvanishing background electric field case.

\section{Introduction of utilizing Matrix Product States to compute the excited states\label{app:mps}}
Matrix product states (MPS)~\cite{ostlund1995thermodynamic}, the entanglement-inspired ansätze for quantum states, represent a paradigmatic family of tensor network states. For systems with $N$ $d$-dimensional sites, the ansätze read as
\begin{equation}
    | \Psi \rangle = \sum_{\{\sigma_i\} \atop \{\alpha_i\}} A^{(0), \sigma_0}_{\alpha_0, \alpha_1} A^{(1), \sigma_1}_{\alpha_1, \alpha_2} \cdots A^{(N-1), \sigma_{N-1}}_{\alpha_{0}, \alpha_N} |\sigma_{0}\sigma_{1}\cdots \sigma_{N-1} \rangle.
\end{equation}
The basis $|\sigma_{0}\sigma_{1}\cdots \sigma_{N-1} \rangle$ is the product state of the local basis $|\sigma_i\rangle$, where $\sigma_i$ runs from $1$ to $d$. $A^{(i), \sigma_i}_{\alpha_{i}, \alpha_{i+1}}$ is a rank-3 parameterized tensor defined on site $i$. For the $i^{\text{th}}$ virtual bond, the summation of bond indices $\alpha_i$ ranges from $1$ to $D_i$, where $D_i$ is called bond dimension and counts the degrees of freedom of the virtual bond connecting $A^{(i-1), \sigma_{i-1}}_{\alpha_{i-1}, \alpha_{i}}$ and $A^{(i), \sigma_i}_{\alpha_{i}, \alpha_{i+1}}$. In the simulation, we usually set $D_i$ as a site-independent number $D$, a controllable value to tune the number of variational parameters. For MPS, $D_0$ is set as $1$ on open boundary conditions and $D$ on periodic boundary conditions. MPS have been proven to be effective ansätze for successfully capturing the information of ground states and low-excited states of one-dimensional gapped systems with local interactions~\cite{hastings2007area,xiang2023density}.

The MPS representing the ground states can be determined by many algorithms based on the imaginary time evolution or the variational optimization methods~\cite{xiang2023density}. Here, we only introduce the variational iterative ground state search since this method is used to obtain the MPS in this paper. For the Hamiltonian $W$, this method minimizes the energy functional
\begin{equation}
    E = \frac{\langle \Psi | W | \Psi \rangle}{\langle \Psi | \Psi \rangle}
\end{equation}
by successively updating local tensors until convergence. Each step extermizes the energy functional with respect to the local tensors to be optimized, which is identical to solving a generalized eigenvalue problem~\cite{schollwock2011density}. In most cases, we need to limit the growing bond dimension of local tensors by performing a low-rank approximation to keep computational costs manageable. This approximation can be achieved using singular value decomposition (SVD) and retaining the particular number of largest singular values. The truncation error, the sum of the squared discarded singular values, measures the difference in the quadratic norm between the original and approximated wave functions.

There are two typical situations when extending this algorithm to determine excited states~\cite{schollwock2011density,li2024accurate}. When the excited state is the ground state of the specific symmetry sector of the Hilbert space, the calculation of the excited state is identical to determining the ground state in that symmetry sector. For the excited states belonging to the same symmetry sector as the ground state, one has two different approaches when computing the MPS representing excited states. 

The first way is to use the purification techniques to target the low-energy subspace density matrix instead of the ground state. After introducing ancillary degrees of freedom, the MPS with ancillary degrees are obtained by applying the imaginary time evolution based algorithms or the optimization based algorithms developed in Ref.~\cite{li2024accurate}. 

Another way is to modify the Hamiltonian by adding a sufficiently large penalty term to shift the calculated low-energy states to highly excited states. Therefore, the excited state next to the shifted states becomes the ground state of the effective Hamiltonian and the algorithms determining the ground state can be reused~\cite{schollwock2011density}. For the ground state $|\Psi_0\rangle$ of the Hamiltonian $W$, a penalty term $\omega_{0} |\Psi_0 \rangle \langle \Psi_0 |$ is added to form the effective Hamiltonian 
\begin{equation}
    W_{\rm{eff}}^1 = W+\omega_0 |\Psi_0 \rangle \langle \Psi_0 |.
    \label{eq:mpsgap}
\end{equation}
When $\omega_0 > E_1 - E_0$ with $E_0$ and $E_i$ being the ground state's energy and the $i$-th excited state's energy, the first excited state $|\Psi_1\rangle$ is exactly the ground state of $W_{\rm{eff}}^1$. Similarly, for $i$-th excited states $|\Psi_i \rangle$, we add the penalty terms $\sum_{k=0}^{i-1} \omega_k |\Psi_k \rangle \langle \Psi_k|$ to form the effective Hamiltonian
\begin{equation}
   W_{\rm{eff}}^{\rm{i}} = W + \sum_{k=0}^{i-1} \omega_k |\Psi_k \rangle \langle \Psi_k|.
\end{equation}
With the constraint $\omega_k > E_{i} - E_{k}$, $|\Psi_i \rangle$ is the ground state of $ W_{\rm{eff}}^{\rm{i}}$. The calculation proceeds sequentially until all desired excited states are obtained. In this paper, we adopt this algorithm to compute the excited states of the Schwinger model as the reference states. The computation is carried out using the open-source ITensors Julia package~\cite{ITensor}. Moreover, the U(1) symmetry is implemented to ensure the zero total charge condition.

\section{Prepare eigenstates on real device by ancillary qubits}
\label{sec:prepare_states_ancillary}
\begin{figure}[!htbp]
    \centering
    \includegraphics[width=0.25\textwidth]{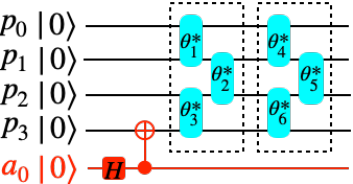}
    \caption{Circuits setup for the preparation of the ground and the first excited state on IBM's device $ibm\_algiers$. The $N=4$ and one ancillary qubit Schwinger model with $x=0.16$, $m/g=0.333$ and $l=0.5$ is taken as a demonstrative case. The $\{ \theta^*_i\}$ are optimized parameters from the simulation. The elements of the Hamiltonian within the subspace can be determined by measuring operator combinations~\cite{xu2023concurrent}. }
    \label{fig:DeviceAncillaResults}
\end{figure}

\renewcommand\arraystretch{1.4}
\begin{table*}[htbp]
    \begin{center}
        \caption{The ancillary qubit results of the ground state (GS) and the first excited state (1st ES) of 4 sites system with $x=0.16$, $m/g=0.333$, and $l=0.5$. For each operator, the results are estimated by 10000 shots. The readout error is mitigated with Twirled Readout Error eXtinction (TREX). The simulation results with the same parameters are given in order to have a direct comparison with the measurement on real device.}
        \setlength{\tabcolsep}{1.55mm}
        \begin{tabular}{c|c|c|c|c|c}
        \hline
        \hline
         Measurement & $ibm\_algiers$ & Simulation & Quantities & $ibm\_algiers$ & ED \\
        \hline
        $H \otimes I_a$ & $1.2723718578 \pm 0.0242834360$ & 1.0062822018 & \multirow{2}*{$E_0$} & \multirow{2}*{0.9983849779} & \multirow{2}*{0.6872150210}\\
        \cline{1-3}
        $H \otimes \sigma_a^x$ & $0.2116411978 \pm 0.0538163233$ & 0.1405063727 & & & \\
        \hline
        $H \otimes  \sigma_a^y$ & $-0.0791128741 \pm 0.0561486229$ & 4.303015$\times {10}^{-17}$ & \multirow{2}*{$E_1$} & \multirow{2}*{1.5463587377} & \multirow{2}*{1.3253490258}\\
        \cline{1-3}
        $H \otimes  \sigma_a^z$ & $0.1549773110 \pm 0.0558908482$ & 0.2864640921 & & & \\
        \hline
        \hline
        $(\sum_n \sigma^z_n) \otimes I_a$ & $0.0136376792 \pm 0.0156884411$ & -3.239860$\times {10}^{-8}$ & \multirow{2}*{$Z_0$} & \multirow{2}*{0.0626140220} & \multirow{2}*{0}\\
        \cline{1-3}
        $(\sum_n \sigma^z_n) \otimes \sigma_a^x$ & $-0.0905303199 \pm 0.0180532985$ & -2.669082$\times {10}^{-8}$ & & & \\
        \hline
        $(\sum_n \sigma^z_n) \otimes \sigma_a^y$ & $-0.0139498889 \pm 0.0208586872$ & -6.051155$\times {10}^{-21}$ & \multirow{2}*{$Z_1$} & \multirow{2}*{-0.0353386635} & \multirow{2}*{0} \\
        \cline{1-3}
        $(\sum_n \sigma^z_n) \otimes \sigma_a^z$ & $0.0299234396 \pm 0.0205232800$ & -1.971089$\times {10}^{-8}$ & &  & \\
        \hline
        \hline
        $(O^2_P/x^2) \otimes I_a$ & $0.9968306147 \pm 0.0134147584$ & 0.9999996212 & \multirow{2}*{$\langle O^2_{P, 0} \rangle/x^2$} & \multirow{2}*{0.3539861313} & \multirow{2}*{0.1335730366}\\
        \cline{1-3}
        $(O^2_P/x^2) \otimes \sigma_a^x$ & $0.7205750237 \pm 0.0162070419$ & 0.8296174093 & & & \\
        \hline
        $(O^2_P/x^2) \otimes  \sigma_a^y$ & $0.0993672448 \pm 0.0170024465$ & -8.398261$\times {10}^{-17}$ & \multirow{2}*{$\langle O^2_{P,1} \rangle/x^2$} & \multirow{2}*{1.6396750982} & \multirow{2}*{1.8663884575}\\
        \cline{1-3}
        $(O^2_P/x^2) \otimes  \sigma_a^z$ & $0.2031834293 \pm 0.0169162521$ & 0.5583316179 & & & \\
        \hline
        \hline
        $L_1 \otimes I_a$ & $0.0491443588 \pm 0.0054756541 $ & 2.517556$\times {10}^{-5}$ & \multirow{2}*{$l_{1,0}$} & \multirow{2}*{0.4143832886} & \multirow{2}*{0.4331668567}\\
        \cline{1-3}
        $L_1 \otimes \sigma_a^x$ & $-0.3941248763 \pm 0.0061581423$ & -0.4149497936 & & & \\
        \hline
        $L_1 \otimes  \sigma_a^y$ & $-0.0534793324 \pm 0.0073345189$ & 2.152435$\times {10}^{-17}$ & \multirow{2}*{$l_{1,1}$} & \multirow{2}*{-0.3160945710} & \multirow{2}*{-0.4331160521}\\
        \cline{1-3}
        $L_1 \otimes  \sigma_a^z$ & $-0.1347843594 \pm 0.0072015880$ & -0.2789114164 & & & \\
        \hline
        \hline
        $(\hat{\Sigma}/g) \otimes I_a$ & $-0.0942649955 \pm 0.0007930895$ & -0.0992034605 & \multirow{2}*{$\Sigma_0/g$} & \multirow{2}*{-0.1721128610} & \multirow{2}*{-0.1852093713}\\
        \cline{1-3}
        $(\hat{\Sigma}/g) \otimes \sigma_a^x$ & $0.0751926081 \pm 0.0009024732$ & 0.0822626211 & & & \\
        \hline
        $(\hat{\Sigma}/g) \otimes  \sigma_a^y$ & $0.0038328134 \pm 0.0010438893$ & -1.279085$\times {10}^{-17}$ & \multirow{2}*{$\Sigma_1/g$} & \multirow{2}*{-0.0164171301} & \multirow{2}*{-0.0131974491}\\
        \cline{1-3}
        $(\hat{\Sigma}/g) \otimes  \sigma_a^z$ & $0.0369000144 \pm 0.0010206520$ & 0.0554453942 & & & \\
        \hline
        \hline
        \end{tabular}
        \label{tab:deviceancillarydata}
    \end{center}
\end{table*}
In this section, we report the results on the hardware for four physical qubits, one ancillary qubit with $x=0.16$, $m/g=0.333$, and $l=0.5$. This demonstration primarily aims to validate the feasibility of preparing eigenstates on near-term quantum devices. Therefore, instead of initiating the cVQE process anew, we choose to employ parameters obtained from the optimized noiseless simulations to directly prepare the ground and first excited states, as Fig.~\ref{fig:DeviceAncillaResults} shows. After preparing the states, subsequent measurements focus on various combinations of operators acting on ancillary and physical qubits to deduce elements pertinent to the target observational quantities. For example, the determination of the Hamiltonian elements within the designated subspace necessitates measuring the initial four operator combinations listed in the "Measurement" section of Table~\ref{tab:deviceancillarydata}. The rotation required to get the eigenstates is ascertained through the diagonalization of the Hamiltonian~\cite{xu2023concurrent}. Additionally, we undertake measurements of operator combinations corresponding to the square of pseudomomentum $O_P^2$, total charge $Z_i$, the electric field on the central link $l_1$, and the chiral condensate $\Sigma_i/g$, as cataloged in the first column of Table~[\ref{tab:deviceancillarydata}].

We demonstrate the feasibility on $ibm\_algiers$. For each operator, we take 10000 shots to estimate the quantities. Additionally, we utilize the Twirled Readout Error eXtinction (TREX) technique to mitigate the readout errors. The measurements with standard error are shown in the second column of Table~[\ref{tab:deviceancillarydata}]. To directly compare the measurements with the ideal expectation values, we also present the noiseless simulation results from the same parameters in the third column of Table~[\ref{tab:deviceancillarydata}]. Utilizing the rotation operator $V$, we extract the physical quantities from our experimental data, which are then enumerated in the fourth column of Table~[\ref{tab:deviceancillarydata}]. The deduced values for the physical quantities associated with both the ground and first excited states are displayed in the fifth column of Table~[\ref{tab:deviceancillarydata}]. Lastly, the sixth column of Table~[\ref{tab:deviceancillarydata}] presents results obtained from ED, enabling a straightforward comparative analysis. The consistency between our measured values and those calculated by ED unambiguously indicates the feasibility of employing the cVQE protocol for eigenstate preparation on actual quantum hardware.

\FloatBarrier
\bibliography{main}

\end{document}